\documentclass[11pt,letterpaper]{article}

\usepackage{graphicx,array}
\usepackage{url}
\usepackage{color}
\usepackage{latexsym}
\usepackage{amsthm}
\usepackage{amsmath}
\usepackage{amssymb}
\usepackage{amsfonts}
\usepackage[numbers,sort&compress]{natbib}
\usepackage{bm}
\usepackage{slashed}
\usepackage{mathrsfs}
\usepackage{enumerate}
\usepackage{tikz}
\usepackage{siunitx}
\usepackage{mdframed}
\usepackage{setspace}  
\usepackage{esvect}

\usepackage{soul}

\usepackage[utf8x]{inputenc}%
\usepackage{tcolorbox}%
\usepackage[normalem]{ulem} 

\usepackage{hyperref} 
\hypersetup{
    colorlinks=true,       
    linkcolor=red,          
    citecolor=blue,        
    filecolor=magenta,      
    urlcolor=blue           
}
\usepackage[all]{hypcap} 

\usepackage{natbib}
\newcommand{\nc}{\newcommand}  

\nc{\beq}{\begin{equation}}  
\nc{\eeq}{\end{equation}}  
\nc{\beqa}{\begin{eqnarray}}  
\nc{\eeqa}{\end{eqnarray}}  
\nc{\bit}{\begin{itemize}}  
\nc{\eit}{\end{itemize}}

\title{
\vspace*{-2.3cm}
\begin{flushright}
\normalsize{
  }
\end{flushright}
\vspace{1.5cm}
\Large  
\textbf{
Diluted Axion Star Collisions with Neutron Stars 
} \vspace*{1.0cm}   
}

\author{{\bf Yang Bai}$^{\,\star}$, {\bf Xiaolong Du}$\,^\diamond$ and {\bf Yuta Hamada}$^{\,\dagger}$
\vspace{5mm}
\\
$^\star$\normalsize\emph{Department of Physics, University of Wisconsin-Madison, Madison, WI 53706, USA}  \vspace{1mm} \\
$^\diamond$\normalsize\emph{Carnegie Observatories, 813 Santa Barbara Street, Pasadena, CA 91101, USA}
\vspace{1mm} \\
$^\dagger$\normalsize\emph{Department of Physics, Harvard University, Cambridge, MA 02138, USA}
}
\date{}

\begin{document}  
\setcounter{page}{0}  
\maketitle  

\vspace*{1cm}  
\begin{abstract} 
Diluted axion star, a self-gravitating object with the quantum pressure balancing gravity, has been predicted in many models with a QCD axion or axion-like particle. It can be formed in the early universe and composes a sizable fraction of dark matter. One could detect the transient radio signals when it passes by a magnetar with the axion particle converted into photon in the magnetic field. Using both numerical and semi-analytic approaches, we simulate the axion star's dynamic evolution and estimate the fraction of axion particles that can have a resonance conversion during such a collision event. We have found that both self-gravity and quantum pressure are not important after the  diluted axion star enters the Roche radius. A free-fall approximate can capture individual particle trajectories very well. With some optimistic cosmological and astrophysical assumptions, the QCD axion parameter space can be probed from detecting such a collision event by radio telescopes. 
\end{abstract} 
  
\thispagestyle{empty}  
\newpage  
  
\setcounter{page}{1}  

\begingroup
\hypersetup{linkcolor=black,linktocpage}
\tableofcontents
\endgroup
\newpage

\vspace{-2cm}

\section{Introduction}
\label{sec:intro}
Axion is a leading candidate for dark matter in the universe~\cite{Turner:1989vc}. Its lightness of mass is protected by an approximated shift symmetry and is cosmologically stable when it is light enough. QCD axion is well motivated because it is a prediction of models solving the strong CP problem (why the neutron electric dipole moment is tiny~\cite{Pendlebury:2015lrz,Graner:2016ses}?) via the Peccei-Quinn symmetry~\cite{Peccei:1977hh,Weinberg:1977ma,Wilczek:1977pj,Shifman:1979if,Kim:1979if,Zhitnitsky:1980tq,Dine:1981rt}.  
Independently, the existence of axion-like particles is ubiquitous in compactifications of string theory~\cite{Svrcek:2006yi,Arvanitaki:2009fg}. In the context of string theory, the axion particle often appears as a pseudo-scalar field from the Kaluza-Klein reduction of the one-form and higher form gauge fields, associated with the non-trivial cycles in the internal geometry. For QCD axion, its couplings to the Standard Model (SM) particles are model-dependent with a non-zero coupling to electric and magnetic fields~\cite{ParticleDataGroup:2020ssz}, which provides the leading interaction for various experimental searches for axion.

If axions are part of dark matter, and if there are sufficiently large fluctuations in the early universe (e.g., spontaneous breaking of the Peccei-Quinn symmetry after inflation), it is known that an object called axion mini-cluster can form~\cite{Hogan:1988mp}. Due to the gravitational cooling effect, some regions of the axion mini-cluster can become colder than other regions by ejecting axion particles, which leads to the formation of axion stars (a self-gravitating system)~\cite{Seidel:1993zk,Levkov:2018kau}. When the axion non-gravitational self-interaction is not important, the formed object is named as the ``diluted axion star", which is a Bose-Einstein condensate (BEC) with gravity balancing the quantum pressure. 

The diluted axion star collisions with neutron stars are phenomenologically interesting.
Since neutron stars produce very strong magnetic fields, photons are expected to be emitted by collisions with axions~\cite{Iwazaki:2014wta,Iwazaki:2015zpb,Raby:2016deh,Bai:2017feq,Amin:2021tnq}.
In particular, if the photon plasma mass around the neutron star is comparable to the mass of the axion, the resonant conversion from axion to photon is efficient~\cite{Huang:2018lxq,Hook:2018iia}. For the QCD axion preferred parameter region, the emitted photons have the frequency of radio waves and could be detected by ground-based telescopes such as the Green Bank Telescope (GBT) and the Square Kilometer Array (SKA).
The resonant conversions of the axion particles for free axion particles and axion particles inside axion mini-clusters are discussed in Refs.~\cite{Huang:2018lxq,Hook:2018iia,Garbrecht:2018akc,Fortin:2021sst,Battye:2021xvt} and \cite{Kavanagh:2020gcy,Edwards:2020afl}, respectively.

On the other hand, there is a large uncertainty on what happens when a diluted axion star passes by a neutron star.~\footnote{The dense axion star does not suffer from the strong tidal force effects~\cite{Buckley:2020fmh,Prabhu:2020yif}. However, the dense QCD axion star~\cite{Braaten:2015eeu} is not stable under the cosmological time~\cite{Visinelli:2017ooc}.} 
 When they are closer than a certain radius (called the Roche radius), the tidal force sourced by the neutron star becomes stronger than the self-gravitational force of the diluted axion star. 
It is possible that the dilute axion star is completely disrupted before reaching the resonance radius where the photon plasma mass becomes the same as the axion mass.
So far, the time evolution of the diluted axion star within the Roche radius is not known. It is the goal of this article to understand the dynamics of a diluted axion star as it collides with a neutron star and to estimate the fraction of axion particles that can have resonant conversion in the magnetic field. 

In detail, we look at the evolution of diluted axion stars under the tidal interaction with a neutron star, which can help to provide a more reliable prediction for the fraction of axion particles that enter the resonance conversion region. In Refs.~\cite{Hui:2016ltb,Du:2018qor}, it is found that for ultralight axions, $m_a\sim 10^{-22}$ eV, the ``tunneling" effect arising from the uncertainty principle or equivalently the quantum pressure leads to an enhanced tidal disruption of dark matter subhalos.
Thus it is interesting to see how this affects the tidal evolution of diluted axion stars which are composed of relatively heavier particles, e.g. QCD axions with a mass around $10^{-5}$~eV.
To this end, numerical simulations are needed. There are several different approaches to simulate the evolution of the axion field. The first approach is
to describe the axions in the non-relativistic limit by a complex wave function satisfying the Schr\"{o}dinger equation \cite{Marsh:2015xka}. Then the Schr\"{o}dinger equation together with the Poisson equation are solved numerically using the finite difference method \cite{Schive:2014dra,Schwabe:2016rze}, or pseudo-spectral method \cite{Woo:2008nn,Mocz:2017wlg,Du:2018qor,May:2021wwp}. This approach captures interesting phenomena related to the wave property of axion particles, e.g. the wave interference,
but requires high computational cost. Another approach is to use the Madelung transform \cite{Madelung:1927} to rewrite the Schr\"{o}dinger equation as quantum fluid Euler equations,
in which there exists an additional pressure term, the so-called quantum pressure, compared to the classical Euler equations. The fluid equation can be solved numerically \cite{Li:2018kyk,Hopkins:2018tht}, or through N-body simulations where the fluid is represented by N-body particles, and the quantum pressure is approximated using particle-in-cell algorithm \cite{Veltmaat:2016rxo} or smoothed-particle hydrodynamics (SPH) \cite{Mocz:2015sda,Nori:2018hud,Nori:2018pka}. This approach is usually much faster, but at the cost of smoothing out part of the
wave behaviors. A hybrid method combining the previous two approaches is also proposed in \cite{Veltmaat:2018dfz}. In this work, we will use the first approach and solve the Schr\"{o}dinger-Poisson equations using pseudo-spectral method where it is possible.
For the cases where the computational cost of pseudo-spectral simulation is too high, we will switch to the SPH approach. At the end, we also introduce the free-particle approximation, where we neglect both the self-gravity and the quantum pressure of the axion star. We cross-check different methods for a fixed $m_a = 10^{-5}\,\mbox{eV}$ to make sure that they
are actually valid in different regimes
\begin{itemize}
\item Axion star with a mass $\gtrsim 10^{-9} M_{\odot}$.~\footnote{Axion star with such a high mass is actually a dense axion star with its stability questionable so far. Anyhow, we will use it to check the SPH simulations by ignoring the non-gravitational self-interaction.} The time scale for the axion star passing the neutron star is comparable to the internal dynamical time scale of the axion star. In this case, the self-gravity and quantum pressure can not be neglected even when the axion star is within the Roche radius. The geometric size of an axion star is relatively small, so we can simulate the evolution of an axion star with sufficient resolution using the pseudo-spectral method;
\item Axion star with a mass in the range $(10^{-12} M_{\odot}, 10^{-9} M_{\odot})$. The self-gravity and quantum pressure are also non-negligible, but their effects are smaller compared to the first case. Because of the large geometric size, the simulation is beyond the capability of the pseudo-spectral method, but the SPH approach still works well;
\item Axion star with a mass $\lesssim 10^{-12} M_{\odot}$. For such a diluted axion star (phenomenologically more likely situation), the internal dynamical time scale is much longer than the axion star passing by time scale, so the self-gravity and quantum pressure can be neglected compared to the gravitational force from the neutron star. In this case, the axion particles are almost free-falling and their trajectories can be solved analytically. 
\end{itemize}

The main difference between the simulations we run in this work and the previous ones, such as the tidal disruption of subhalo cores (the axion star is at the center of  a subhalo) \cite{Du:2018qor}, is that the axion stars considered in this work are not gravitationally bounded to a neutron star, so the time they interact with a neutron star is much shorter. Even when axion particles are no longer bounded to the axion star, they are still on a similar orbit. But in the latter case, the axion star (or the subhalo core) orbits the host halo and only gradually sinks to the center of the host halo due to dynamical friction. Thus the axion star has sufficiently long time to lose its mass and become totally disrupted.

On the phenomenological side, we also study the possibility that radio waves caused by the collision of axion and neutron stars can be detected by the GBT and SKA.
We consider the collision events in the Milky Way, Andromeda galaxy (M31), and the globular cluster M54 at the center of the nearby Sagittarius dwarf galaxy.
We find that, up to cosmological and astrophysical uncertainties (for instance, the fraction of axion stars in the total dark matter density and the dark matter profile), the sensitivity can reach the QCD axion parameter region.

The structure of the paper is as follows. In Section~\ref{sec:axion-stars}, we review the basic properties of the dilute axion star. In Section~\ref{sec:trajectory}, we study the tidal evolution of an axion star passing a neutron star through numerical and semi-analytic methods, and compute the fraction of axion particles that can have resonance conversion. In Section~\ref{sec:pheno}, we investigate the possibility to detect the radio wave signal for such an encounter event by radio telescopes. The discussion and conclusions are presented in Section~\ref{sec:conclusion}. The technical details are given in the Appendix~\ref{sec:ps_method} and Appendix~\ref{sec:approximation}.

\section{Diluted axion star}
\label{sec:axion-stars}
Starting with the relevant interactions for the axion particle
\beqa
{\cal L} \supset \, -\frac{1}{4}F_{\mu\nu}F^{\mu\nu}\, +\, \frac{1}{2}(\partial_\mu a)^2 \,- \, \frac{m_a^2}{2} a^2 
- \frac{g_{a\gamma\gamma}}{4}a\,F_{\mu\nu}\widetilde{F}^{\mu\nu} 
- c_\psi \, \frac{\partial_\mu a}{f_a} \,\overline{\psi} \gamma^\mu \gamma^5 \psi ~,
\eeqa
where $- \frac{g_{a\gamma\gamma}}{4}a\,F_{\mu\nu}\widetilde{F}^{\mu\nu} = g_{a\gamma\gamma}\,a\,\mathbf{E}\cdot\mathbf{B}$, $g_{a\gamma\gamma} = \frac{\alpha_{\rm em}}{2\pi\,f_a}(E/N - 1.92) = (0.203\,E/N - 0.39)\,m_a/\mbox{GeV}^2$ with $\alpha_{\rm em}\approx 1/137$ and $E$ and $N$ related to the electromagnetic and QCD anomalies of the current associated with the axion~\cite{ParticleDataGroup:2020ssz}. $E/N=8/3$ for DFSZ~\cite{Dine:1981rt,Zhitnitsky:1980tq} models and $E/N=0$ for KZVZ~\cite{Kim:1979if,Shifman:1979if} models.
Here, $\widetilde{F}^{\mu\nu} = \frac{1}{2}\epsilon^{\mu\nu\rho\sigma}F_{\rho\sigma}$ and $\psi$ represents SM fermions. Due to the smallness of the axion mass, its occupation number can be very large such that one can use a wave description for the axion particles or fields. Since axions are non-relativistic inside a diluted axion star,~\footnote{When axion stars passing through a neutron star, only a small fraction of axion particles will have a large relativistic velocity, so we will ignore the relativistic effects in this study.} one can perform a non-relativistic expansion $a(\mathbf{x}, t) = [\psi(\mathbf{x}) e^{-im_a  t} + \psi^*(\mathbf{x}) e^{i m_a t}]/\sqrt{2m_a}$ and derive the Schr\"odinger and Poisson equations. Throughout this paper, we will use the natural unit with $c=1$ and $\hbar = 1$.
\beqa
i\,\frac{\partial \psi}{\partial t} &=& - \frac{1}{2m_a}\nabla^2\psi + m_a\,\Phi_{\rm total}(\mathbf{x}, t) \,\psi  \,,\ 
\label{eq:schr} \\
\nabla^2 \Phi_{\rm self} &=& 4\pi \, G_N\, m_a\, |\psi|^2 ~.
\label{eq:poisson}
\eeqa
Here, $G_N$ is the Newton constant. The total gravitational potential $\Phi_{\rm total} = \Phi_{\rm self} + \Phi_{\rm NS}$ with $\Phi_{\rm self}$ as the axion star self-gravitational potential and $\Phi_{\rm NS}$ as the external one provided by a neutron star. When the axion star is far away from the neutron star, one could ignore $\Phi_{\rm NS}$, while when it is very close to the neutron star, one could ignore $\Phi_{\rm self}$. 

Treating the axion as a fluid, one has 
\beqa
\psi\equiv \sqrt{\frac{\rho}{m_a}}\,e^{i\theta} \;,\quad
\mathbf{v} \equiv \frac{1}{m_a}\bm{\nabla}\theta = 
\frac{1}{2i\,m_a}\left( \frac{1}{\psi}\,\boldsymbol{\nabla}\psi - \frac{1}{\psi^*}\,\boldsymbol{\nabla}\psi^* \right)~.
\eeqa
The continuity and Euler equations are
\beqa
\dot{\rho} + \bm{\nabla} \cdot (\rho\,\mathbf{v}) &= & 0 ~, \label{eq:continuity}\\
\dot{\mathbf{v}} + (\mathbf{v}\cdot \bm{\nabla}) \mathbf{v} &=& - \bm{\nabla} \, \Phi_{\rm total} + \frac{1}{2\,m_a^2} \, \bm{\nabla} \left( \frac{\nabla^2\sqrt{\rho}}{\sqrt{\rho}} \right) ~. 
\label{eq:Euler}
\eeqa
The last term of the above equation can be treated as the quantum pressure of the system: it costs energy to have an over-densed region. 

For an isolated axion star with $\Phi_{\rm total} = \Phi_{\rm self}$, one can solve the stationary Schr\"odinger-Poisson system to obtain the profiles of the axion star. For the spherically-symmetric ground state, the profile can be fitted well by the following power-law behavior~\cite{Schive:2014dra,Marsh:2015wka},
\beqa
\rho(r) = \frac{\rho_c}{[1 + \alpha^2\,(r/R_c)^2]^8} ~, 
\label{eq:rho_star}
\eeqa
with $\alpha \approx 0.3$ and $\rho_c$ as the central density. The parameter $R_{c}$ is the radius where the density drops to $1/2$ of the central density. The axion star radius is defined to be the 99\%-enclosed-mass radius as $R_{\rm AS} \approx 3.9 R_c$. For a given axion particle mass and a axion star mass, one has~\cite{Hui:2016ltb}
\beqa
\label{eq:axion-star-radius}
R_{\rm AS} = \frac{10.5}{G_N \, M_{a\odot}\,m_a^2} = (2.7\times 10^{4}\,\mbox{km})
\left( \frac{10^{-15}\,M_{\odot}}{M_{a\odot}} \right)\,\left( \frac{10^{-4}~\mbox{eV}}{m_a} \right)^2 ~. 
\eeqa
For the axion stars in the diluted axion star branch, the axion star mass is bounded from above or~\cite{Chavanis_2011,Braaten:2015eeu}
\beqa
\label{eq:diluted-mass-condition}
M_{a\odot} \lesssim 10.15\, \frac{f_a}{m_a\,G_N^{1/2}} \approx (1.1\times 10^{-14}\,M_\odot)\,
\left( \frac{10^{-4}\,\mbox{eV}}{m_a} \right) \, \left( \frac{f_a}{10^{11}\,\mbox{GeV}} \right)  ~.
\eeqa
%

\section{Tidal evolution of an axion star passing a neutron star}
\label{sec:trajectory}
We discuss the dynamical evolution of an axion star around a neutron star when the axion star is inside the Roche radius $R_{\rm Roche}$ (the radius where the self-gravitational force equals to the tidal force). The Roche radius is estimated as (see e.g. \cite{Buckley:2020fmh})
\beq
R_{\rm Roche}:=R_{\rm  AS}\left(\frac{2\,M_{\rm NS}}{M_{a\odot}}\right)^{1/3}
\simeq
\left(1.26\times10^6\,\mbox{km}\right) \left(\frac{R_{\rm AS}}{10^2\,\mbox{km}}\right)\left(\frac{M_{\rm NS}}{M_\odot}\frac{10^{-12}M_\odot}{M_{a\odot}}\right)^{1/3} ~.
\eeq

The density evolution of the axion star is determined by three effects: (1) tidal force from the neutron star; (2) self-gravity; (3) quantum pressure. Without the external tidal force, the self-gravity is balanced by the quantum pressure, leading to a stable solitonic solution. Under the tidal force from the neutron star, the axion star will deform and be gradually disrupted. It is shown in \cite{Hui:2016ltb,Du:2018qor} that the quantum pressure may play an important role by enhancing the tidal mass loss. However, in this paper we mainly concentrated on head-on collisions, in which case the passing time of the axion star through the neutron star is very short, so the tidal mass loss is not significant within the time scales we are concerned with.

To estimate how important each of the three effects mentioned above is, we compare several time scales below.
\begin{itemize}
\item
The axion star crossing time scale: the time taken for the axion star going from $R_{\rm Roche}$ to the surface of the neutron star. As an order of magnitude estimation, it has
\begin{equation}
T_{\rm cross} \approx \frac{ R_{\rm Roche}}{\sqrt{2\, G_N\, M_{\rm NS}/R_{\rm Roche}+v_{\rm i}^2}} ~,
\label{eq:T_cross}
\end{equation}
where $v_{\rm i}$ is the velocity of axion star at infinity.
\item
The dynamical time scale of the axion star:
\begin{equation}
T_{\rm dyn} = \sqrt{\frac{R_{\rm AS}^3}{G_N\, M_{a\odot}}} ~.
\label{eq:T_dyn}
\end{equation}
This is also the time scale that the axion star responses to the external perturbations. For example, Ref.~\cite{Guzman:2004wj} shows that
the perturbed axion star oscillates with a period of roughly $2.7\,T_{\rm dyn}$.
\end{itemize}

\begin{figure}[t]
\begin{center}
\includegraphics[width=.6\textwidth]{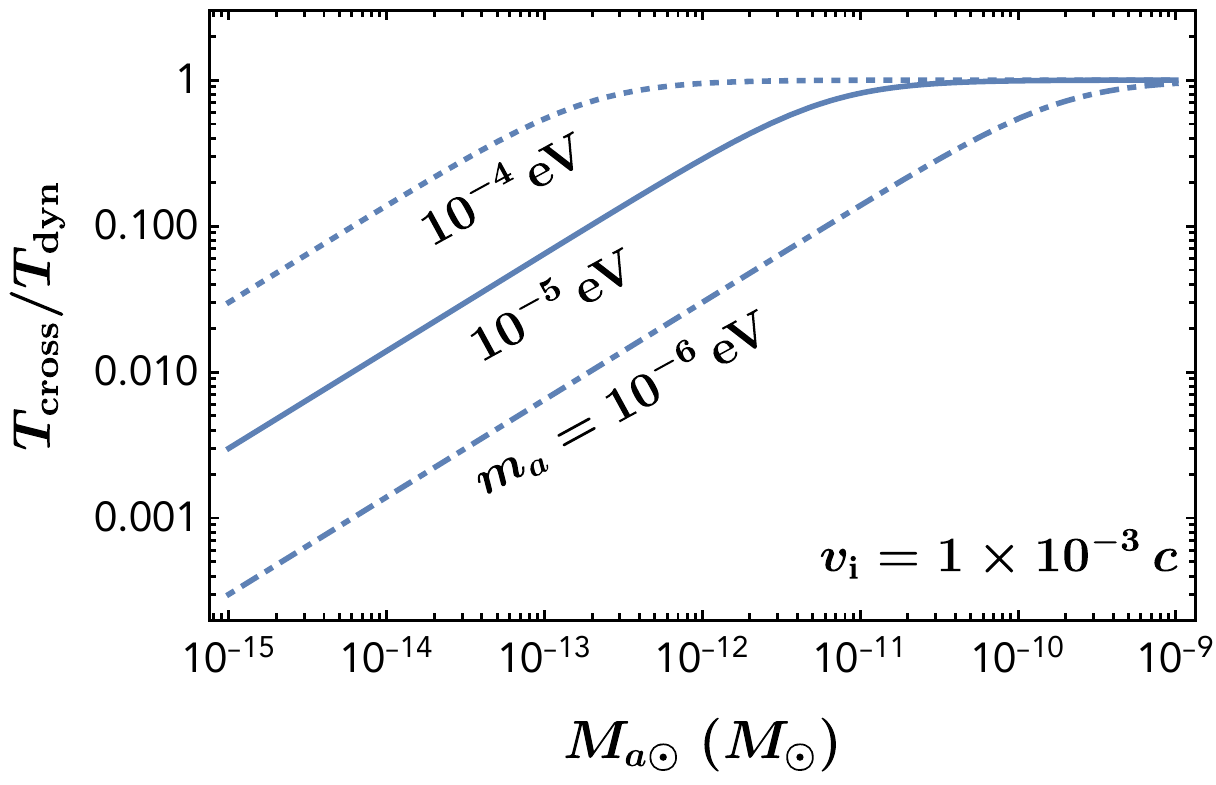}
\end{center}
\caption{The ratio of the crossing time over the dynamical time for different axion star masses. The initial velocity is chosen to be $v_{\rm i} = 1\times 10^{-3}\,c$.}
\label{fig:timeratio}
\end{figure}

In Fig.~\ref{fig:timeratio}, we plot the ratio of $T_{\rm cross}$ over $T_{\rm dyn}$ as a function of the axion star mass.
In most cases we consider, i.e. diluted axion stars, the passing time scale is shorter than the dynamical time scale of the axion star. So the particles in the axion star do not have enough time to be re-virialized. In other words, the density evolution is dominated by the tidal force. Noticing this fact can dramatically simplify the calculation to obtain the whole (approximate) evolution for the axion star passing the neutron star.

\subsection{Results from wave simulations}
To see how the axion star evolves during the collision, we simulate the head-on collision of an axion star with a neutron star by solving the Schr\"{o}dinger-Poisson equations (\ref{eq:schr}) and (\ref{eq:poisson}) using pseudospectral method as introduced in Ref.~\cite{Du:2018qor}. Since the system has a rotational symmetry with respect to the axis joining the axion star and neutron star, we can simplify the problem to a two-dimensional problem and solve it in the cylindrical coordinates $(\eta,\varphi,z)$ where $\eta$ is the axial distance, $\varphi$ is the azimuth, and $z$ is the height. Due to the rotational symmetry, the wave function only depends on $\eta$ and $z$. We have changed the way to discretize the spatial space that used by \cite{Du:2018qor}. More specifically, we discretize the $z$ coordinate using uniform grid, but use non-uniform Chebyshev grid for $\eta$ coordinate. Accordingly we impose periodic boundary conditions in the $z$ direction, but zero boundary conditions in the $\eta$ direction.
Note that $\eta\geq0$, but we have extended it to negative values by assuming $\psi(-\eta,z)=\psi(\eta,z)$. In practice, the simulation box is chosen to be sufficiently large so that the boundary effect does not affect the wave function in the central region we are concerned with. We have checked that our two-dimensional simulation gives consistent results as that from  a three-dimensional simulation using the same method as in \cite{Du:2018qor}. More details about the numerical algorithm can be found in Appendix \ref{sec:ps_method}.

At the initial time, the wave function is set up from the density profile of the axion star, Eq. (\ref{eq:rho_star}), and the initial distance
between the axion star and neutron star is chosen to be far enough so that the shape of axion star is not significantly affected by the gravity of the neutron star yet.
We work in the frame centered on the axion star and model the gravity from the neutron star by adding an external tidal potential that satisfies
\begin{eqnarray}
\nabla \Phi_{\rm tital}(\mathbf{x})&=&\nabla\Phi_{\rm NS}(\mathbf{x})-\nabla\Phi_{\rm NS}(0) ~,
\label{eq:Phi_t_1}
\\
\Phi_{\rm tital}(0)&=&0 ~,
\label{eq:Phi_t_2}
\end{eqnarray}
where $\Phi_{\rm NS}$ the gravitational potential from the neutron star. To the lowest order with respect to $|\mathbf{x}|/R$, where $R$ is the distance to the neutron star, the tidal potential is proportional to $|\mathbf{x}|^2$. The neutron star is modeled as a sphere with a uniform density. The relative position of the neutron star at time $t$ is set by solving the motion of a test particle in the gravitational potential of the neutron star.

Due to computational limitation, we simulate a relative compact axion star with $M_{a\odot}=10^{-9} M_{\odot}$ for $m_a=10^{-5}$ eV. The axion star radius is $R_{\rm AS} \approx 2.7\,\mbox{km}$ from \eqref{eq:axion-star-radius}. The neutron star has a mass of $1\,M_{\odot}$ and a radius of $10$ km. At the initial time, the axion star is placed at a relative distance of
$D_{\rm ini}=2958\,{\rm km}\sim R_{\rm Roche}$ and has an initial relative velocity of $v_{0}=0.03\,c$ (velocity at infinity $v_{\rm i} = 10^{-3}\,c$). With this set-up, $T_{\rm cross} \approx T_{\rm dyn} \approx 0.385\,{\rm s}$.
Note that the actual dynamical time scale is radius-dependent, i.e. $T_{\rm dyn}$ is smaller at smaller radii where the density is higher. Thus the innermost part will be affected most by the self-gravity and quantum pressure. The axion star is evolved till it reaches a relative distance of $\sim 85\,{\rm km}$ when most of the axion particles are already within the the resonance radius $R_{\rm res} \sim 100\,{\rm km}$ from the neutron star. The simulation is run in a rectangular box with $\eta \in(-5.3,+5.3)\,\rm{km}$ and $z\in(-17.7,+17.7)\,\rm{km}$. Considering that the tidal acceleration and internal motion of axion particles induced by the tidal force in the $z$ direction are much larger, we use a higher resolution for $z$ with $16384$ grid points, while for $\eta$ we use a lower resolution with $512$ grid points. We have checked that the resolution is sufficient till the end of the simulation (see Eq. [\ref{eq:x_res}]).

\begin{figure}[t!]
\begin{center}
    \includegraphics[width=.9\textwidth]{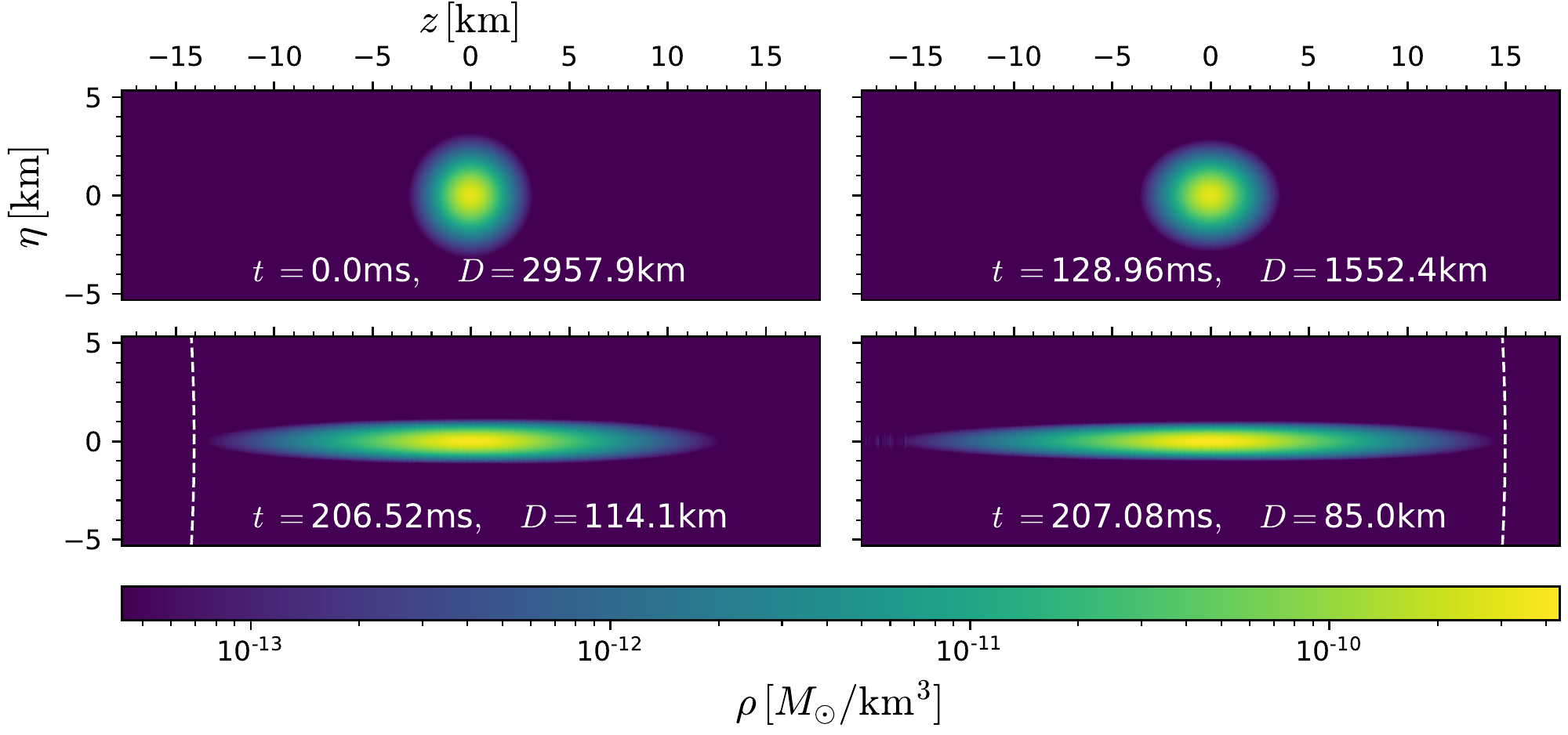}
\end{center}
\caption{Head-on collision of an axion star of mass $M_{a\odot}=10^{-9} \,M_{\odot}$ with a neutron star of mass $1\,M_{\odot}$. The axion
particle mass $m_a=10^{-5}\,\mbox{eV}$. Different panels show the slice density of the axion star across its center at different times from the pseudospectral simulation. The initial velocity of axion star is set by assuming that it has a velocity of $10^{-3}\,c$ at infinity. The white dashed lines in the lower panels indicate the $100\, {\rm km}$ radius (the resonance conversion region) with respect to the neutron star.}
\label{fig:slice}
\end{figure}

Figure~\ref{fig:slice} shows the slice density of the axion star across its center at different simulation times. The axion star moves in the $-z$ direction. At the initial time (top-left panel), the axion star is assumed to be spherically symmetric. As it approaches the neutron star, it gradually deforms under the tidal force, i.e. stretched in the $z$ direction and compressed in the $\eta$ direction (top-right panel). When the
axion star reaches the $100\,{\rm km}$ radius (white dashed line in the lower panels) from the neutron star, its radius in the moving direction is already several times larger than its initial radius (lower-left panel). So it will take longer for all the axion articles to pass through the $100\,{\rm km}$ radius than the case assuming unchanged shape of axion star. This effect will be more significant for more diluted axion stars considered in the following sections. As most of the axion particles have past through the $100\,{\rm km}$ radius, the axion star is further stretched in the moving direction and its shape becomes slightly non-symmetric on the near side and far side with respect to the neutron star (lower-right panel). At an even later time, the internal velocities of axion particles relative to the center of axion star become very large and are beyond the spatial resolution. So we will rely on smoothed-particle simulations and semi-analytic calculations introduced in the following two subsections.

\subsection{Results from smoothed-particle hydrodynamics simulations}

While the pseudospectral method solves the full dynamics of axion stars, its application is limited by its high requirement for computational
resources. To resolve the high velocity field when the axion star is close to the neutron star, e.g. at $100\,{\rm km}$ distance, an extremely high
spatial resolution is required
\begin{equation}
\Delta z < \frac{\pi}{m_a v_{z, \rm max}}\,,\quad \Delta \eta < \frac{\pi}{m_a v_{\rm \eta, max}} \,,
\label{eq:x_res}
\end{equation}
where $v_{z, {\rm max}}$, $v_{\eta, {\rm max}}$ are the maximum velocity of axion field in the center-of-mass coordinates along the $z$ and $\eta$ directions, respectively. Due to the tidal force, axion particles in the axion star can acquire a very large velocity with respect to the center of mass, making it difficult to simulate the evolution of axion star when it enters the
resonance conversion region, especially for more diluted axion stars which have much
larger radii (a larger simulation box is needed). Using the pseudospectral method described in the previous section, we are only
able to simulate an axion star of mass $10^{-9} M_{\odot}$ for $m_a=10^{-5}\,{\rm eV}$, which is far away from the more interesting
region with $M_{a\odot}\sim10^{-15} M_{\odot}$.

To push the numerical simulation further into the region we would like to investigate, we employ another approach, the smoothed-particle
hydrodynamics (SPH) simulations. In this approach, an additional acceleration is added to the N-body particles to account for the quantum
pressure (see Eq. [\ref{eq:Euler}]). This method has been used to simulate the structure formation in ultralight axion dark matter
models and shows high efficiency \cite{Mocz:2015sda,Nori:2018hud,Nori:2018pka}.~\footnote{Note that the interference phenomenon is not
well resolved by the SPH approach due to the smoothing operations, but it is irrelevant for the study in this work.}

In the SPH framework, an arbitrary function $A(\mathbf{x})$ is approximated by
\begin{equation}
A(\mathbf{x_i}) = \int A(\mathbf{x})\,W(\mathbf{x_i}-\mathbf{x},h_i) d^3 \mathbf{x} \approx \sum_j \frac{m_j}{\rho_j} \, A(\mathbf{x_j})\,
 W_{ij} ~,
\label{eq:sph_A}
\end{equation}
where $\mathbf{x_i}$ and $\mathbf{x_j}$ are the positions of the $i$-th and $j$-th particles, $m_j$ is the mass of the particle $j$,
and $W_{ij}\equiv W(\mathbf{x_i}-\mathbf{x_j},h_i)$ is a smoothing kernel function. The density at the position of particle $i$ is given by
\begin{equation}
\rho_i =  \sum_j m_j W_{ij} ~.
\label{eq:sph_rho}
\end{equation}
Usually we also need to calculate the derivative of the field $A(\mathbf{x})$. This can be done by simply taking derivatives
of both sides of Eq. (\ref{eq:sph_A})
\begin{equation}
\nabla A(\mathbf{x_i}) \approx \sum_j \frac{m_j}{\rho_j}\, A(\mathbf{x_j})\, \nabla W_{ij} ~.
\label{eq:sph_A_der}
\end{equation}
Similarly, the second derivative is approximated by
\begin{equation}
\partial^2_{x,y} A(\mathbf{x_i}) \approx \sum_j \frac{m_j}{\rho_j} \,A(\mathbf{x_j}) \,\partial^2_{x,y} W_{ij} ~.
\label{eq:sph_A_der2}
\end{equation}
With these formula, we can calculate the values of a field and its derivative at the particle positions.

To calculate the acceleration from quantum pressure, we follow the approach by Ref.~\cite{Mocz:2015sda} and introduce
an effective pressure tensor
\begin{equation}
P^Q_{x^{\alpha} x^{\beta}} = \frac{1}{4}\frac{1}{m_a^2}\left(\frac{1}{\rho}\frac{\partial \rho}{\partial x^{\alpha}}\frac{\partial \rho}{\partial x^{\beta}}
-\frac{\partial^2\rho}{\partial x^{\alpha} \partial x^{\beta}}\right) ~,
\label{eq:sph_PQ_2}
\end{equation}
so that the last term of Eq. (\ref{eq:Euler}) can be written as
\begin{equation}
\frac{1}{2\,m_a^2} \, \bm{\nabla} \left( \frac{\nabla^2\sqrt{\rho}}{\sqrt{\rho}} \right)
= - \frac{1}{\rho}\nabla\cdot{\mathbf{P^Q}} ~.
\label{eq:sph_PQ}
\end{equation}
To be more accurate, we also include correction terms when calculating the first and second derivatives as suggested by \cite{Nori:2018hud}.

We have implemented the above calculations in the public N-body/SPH code, {\small{GADGET}}-4~\cite{Springel:2020plp}. To check how well this approach
works, we run a simulation with the same initial condition used in the previous section and compare the results from these two methods. We find excellent agreements between these two approaches (see Fig.~\ref{fig:comparison}).

\begin{figure}[t!]
\begin{center}
    \includegraphics[width=.9\textwidth]{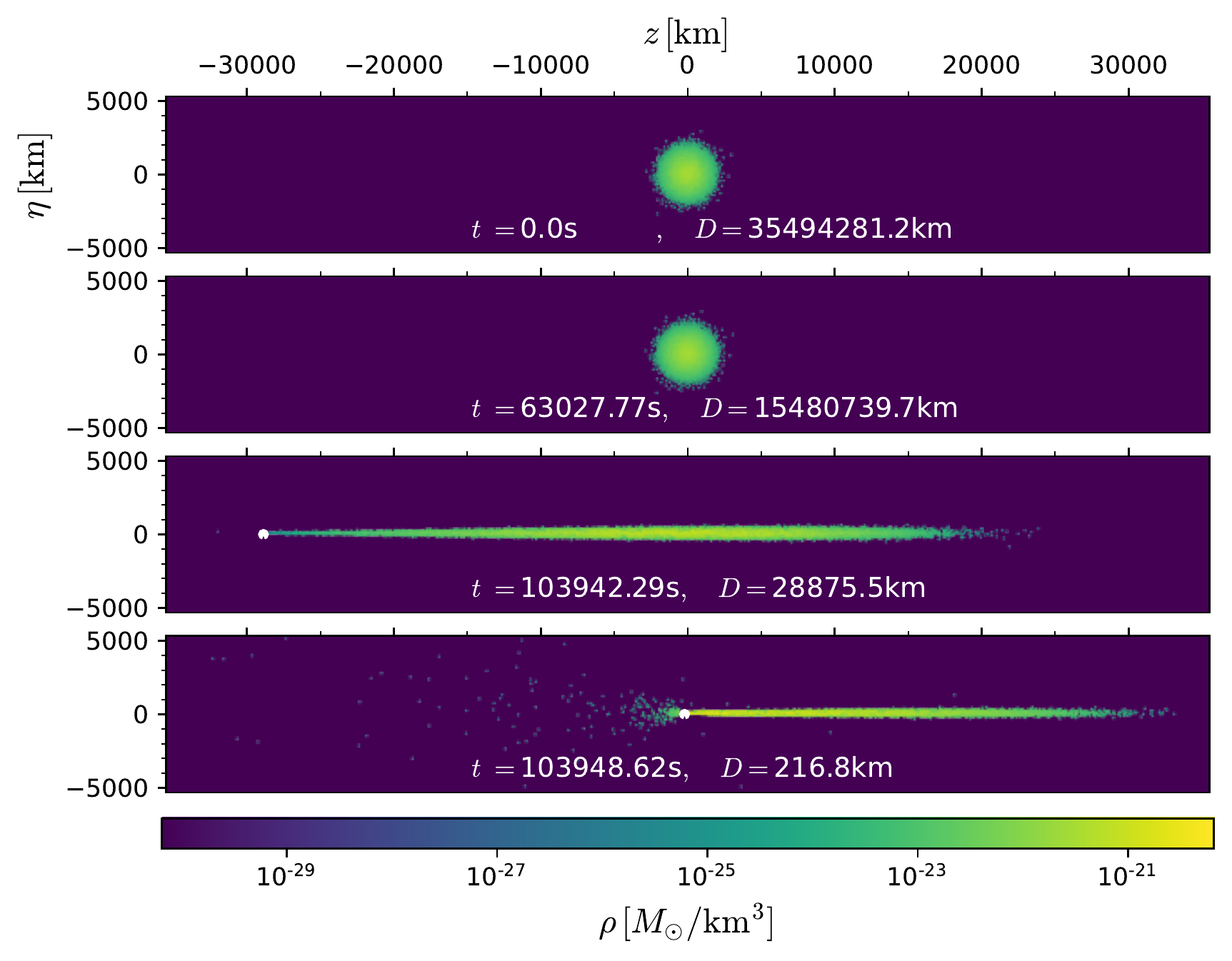}
\end{center}
\caption{Head-on collision of an axion star of mass $M_{a\odot}=10^{-12} \,M_{\odot}$ with a neutron star of mass $1\,M_{\odot}$. The axion
particle mass $m_a=10^{-5}\,\mbox{eV}$. Different panels show the slice density of the axion star across its center at different times from the SPH simulation. The initial velocity of axion star is set by assuming it has a velocity of $10^{-3}\,c$ at infinity. The small white circles in the lower panels indicate the $100\, {\rm km}$ radius with respect to the neutron star.}
\label{fig:slice_M12}
\end{figure}

Using the SPH code, we run another simulation of a more diluted axion star with $M_{a\odot}=10^{-12} \,M_{\odot}$, $m_a=10^{-5}\,{\rm eV}$. The slice density at different times is shown in Fig.~\ref{fig:slice_M12}. In this case, the axion star radius is $2691$ km, i.e. $10^3$ times of the case discussed in the previous section. The results are qualitatively similar to the more compact one. But in this case, the crossing time scale relative to the dynamical time scale of the axion is shorter, so the quantum pressure and self-gravity plays a less important role. At the time when the tidal radius equals to the half-mass radius of the axion star (panel 2), we see little change in the shape (cf. top-right panel in Fig.~\ref{fig:slice}). When the front end of the axion star enter $100$ km radius (panel 3), the axion star is significantly stretched along $z$ direction, the size is almost $10$ times of the initial value. So it takes longer time for whole axion star to pass through the neutron star, generating a radio signal with a longer duration. In the $\eta$ direction, the axion star is compressed making more axion particles be able to enter the resonance conversion radius. When the axion star passes through the neutron star, it is destroyed by the strong tidal force from the neutron star (panel 4).

\begin{figure}[th!]
\begin{center}
    \includegraphics[width=.8\textwidth]{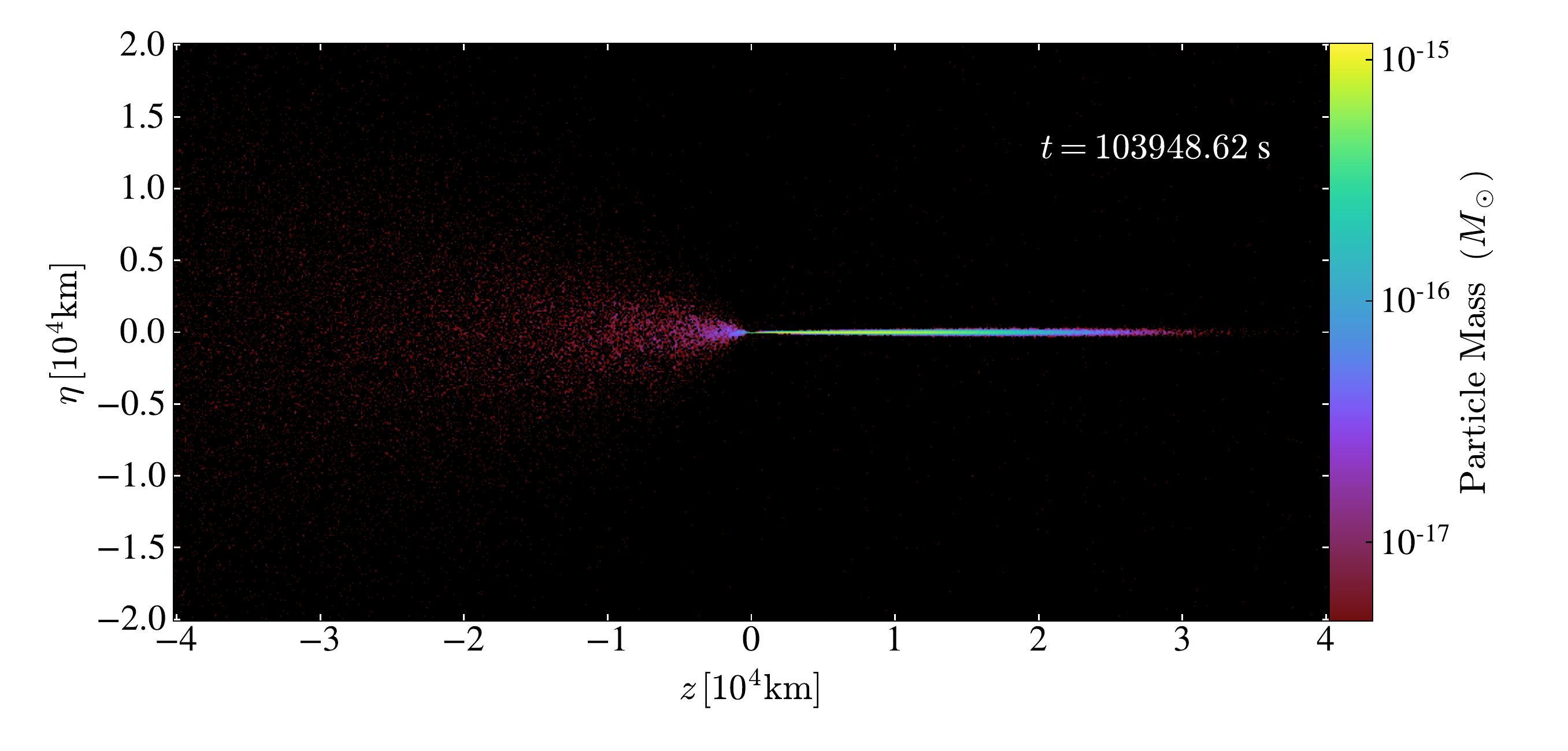}
    \includegraphics[width=.8\textwidth]{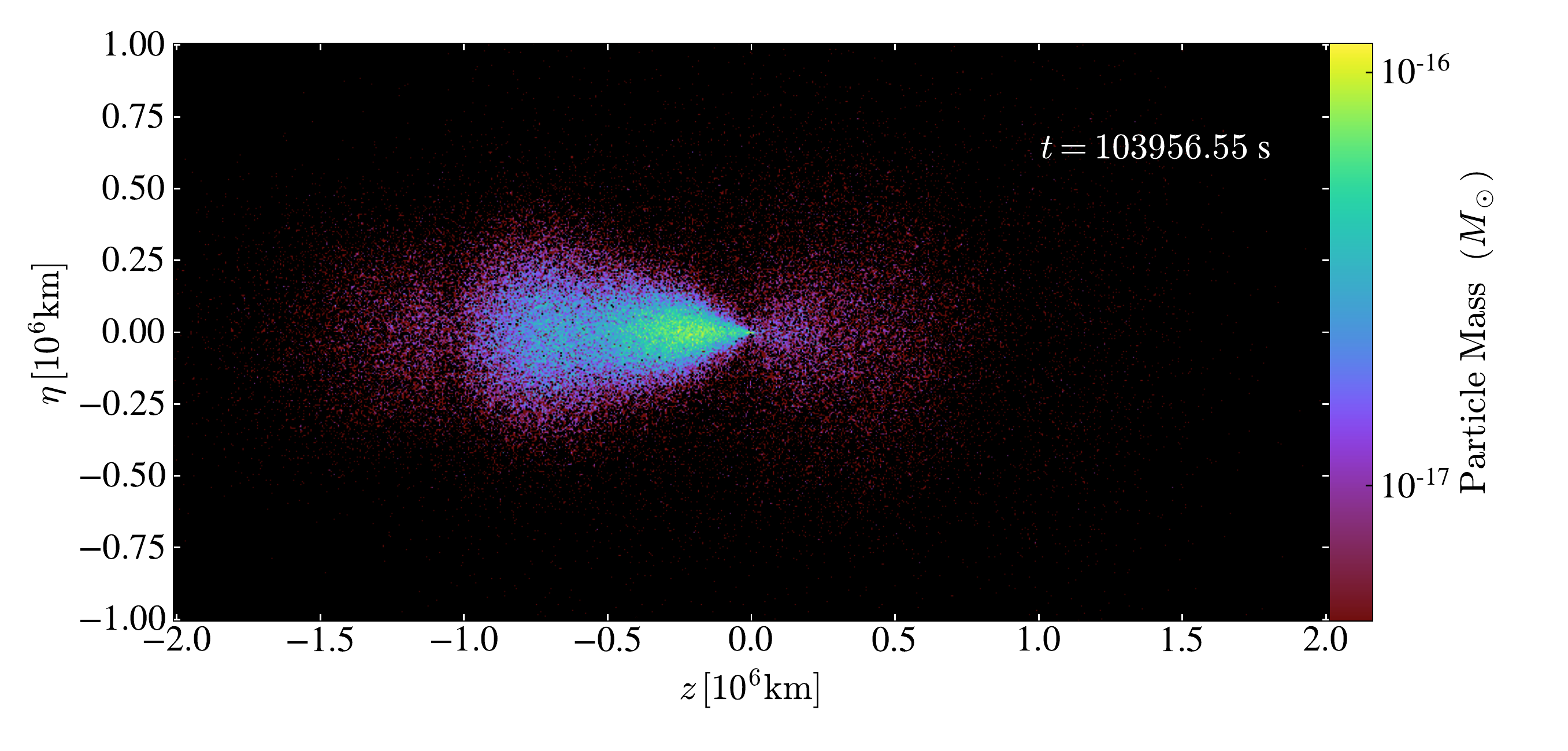}
\end{center}
\caption{Projected (after integration over one direction in the Cartesian coordinates) particle distributions from the same simulation as in Fig.~\ref{fig:slice_M12}. The color encodes the projected particle mass in each pixel. Note that the coordinate system is centered on the neutron star. The upper panel corresponds to the bottom panel in Fig.~\ref{fig:slice_M12}. The lower panel shows the results at the time when most of axion particles have past through the neutron star.
}
\label{fig:slice_M12_proj}
\end{figure}

Figure \ref{fig:slice_M12_proj} shows the projected density of the axion star when half and almost all the axion particles have past through the neutron star. At the end, we see that axion star is totally disrupted with the particles distributed over a large space volume. A small fraction of axion particles are reflected back to the incoming direction.

\subsection{Results from free-fall particle approximation}

The calculation is simplified if we assume the self-gravity and quantum pressure of the axion star are neglected.
In fact, we can obtain the (semi-)analytic formula of the trajectory of the particles (see Appendix~\ref{sec:approximation}).
We call this the free-fall particle approximation.

Before the axion star reaches the Roche radius, the self-gravity balances with the quantum pressure, and we do not expect that the free-fall particle approximation is valid. 
On the other hand, inside the Roche radius, the external gravitational force sourced by the neutron star dominates over the others, and we can treat the axion star as the collection of the free-fall particles.
In the following, we will show that this is indeed the case by comparing the numerical simulation and semi-analytic calculation based on the free-fall particle approximation.

In order to provide the justification of the free-fall particle approximation, we compare it with the wave and SPH simulations for $m_a=10^{-5}\,{\rm eV}$ and $M_{a\odot}=10^{-9}\,M_\odot \, ,10^{-12}\,M_\odot$.
We use the same parameters as in Figs. \ref{fig:slice} and \ref{fig:slice_M12} and start the calculations from top-right panel in Fig.~\ref{fig:slice} and the second panel in Fig.~\ref{fig:slice_M12}, respectively. 
At this time, the tidal radius equals to the half-mass radius of the axion star. The density and internal velocity are well fitted by values in Table \ref{Tab:initial_condition} with the parametrization density function of
\begin{align}
&\rho(z,\eta) = \frac{\rho_c}{\left[1 + \alpha^2\,\left(\frac{\eta^2}{\eta_c^2} + \frac{z^2}{z_c^2}\right)\right]^8} ~.
\label{eq:fitting_density}
\end{align}
\begin{table}[hb!]
  \renewcommand{\arraystretch}{1.5}
    \addtolength{\tabcolsep}{3pt} 
\begin{center}
\small
\begin{tabular}{c|ccccc}\hline \hline
  Mass    &  $\rho_c\,[M_\odot/\text{km}^3]$      & $\eta_c\,[\text{km}]$   & $z_c\,[\text{km}]$ & $v_\eta\,[\text{km}/\text{s}]$    &  $v_z\,[\text{km}/\text{s}]$  \\\hline\hline
$M_{a\odot}=10^{-9}\,M_\odot$   &  $3\times10^{-10}$   &  $0.77$  
& $0.61$ & $-1.90 \left(\eta/\text{km}\right)$ &  $3.23 \left(z/\text{km}\right)$  \\\hline
$M_{a\odot}=10^{-12}\,M_\odot$   &  $3\times10^{-22}$ &  $662.662$   & $686.208$ & $-7.10\times10^{-7}  \left(\eta/\text{km}\right)$
& $1.38\times10^{-6}\left(z/\text{km}\right)$ \\\hline
\end{tabular}
\caption{Shape and internal velocity of the axion star when the tidal radius equals to the half-mass radius (top-right panel in Fig.~\ref{fig:slice} or second panel in Fig.~\ref{fig:slice_M12}). Here $v_\eta$ and $v_z$ are the $\eta$ and $z$ components of the velocity field in the axion star center of mass frame, respectively.
}
\label{Tab:initial_condition}
\end{center}
\end{table}

\begin{figure}[th!]
 \begin{center}
  \includegraphics[width=.48\textwidth]{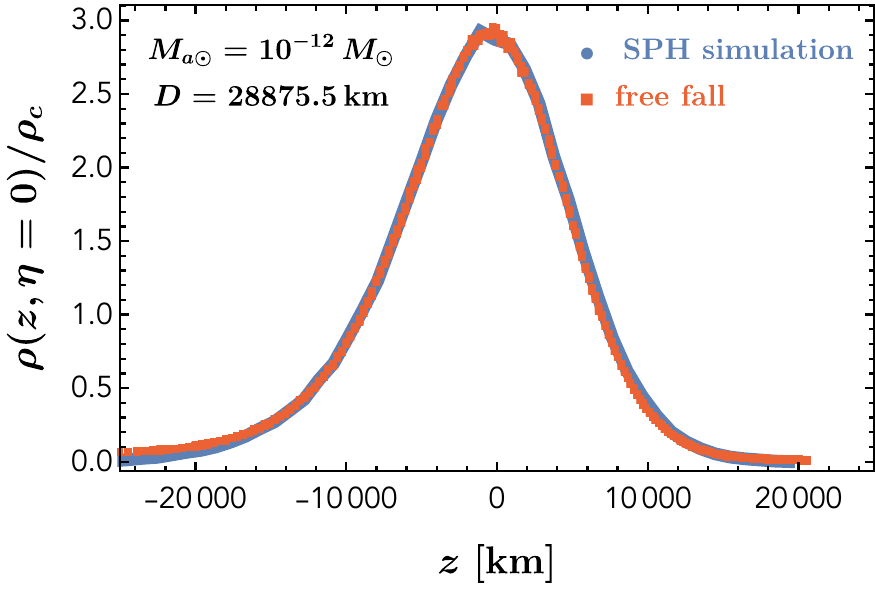} \hspace{3mm}
  \includegraphics[width=.48\textwidth]{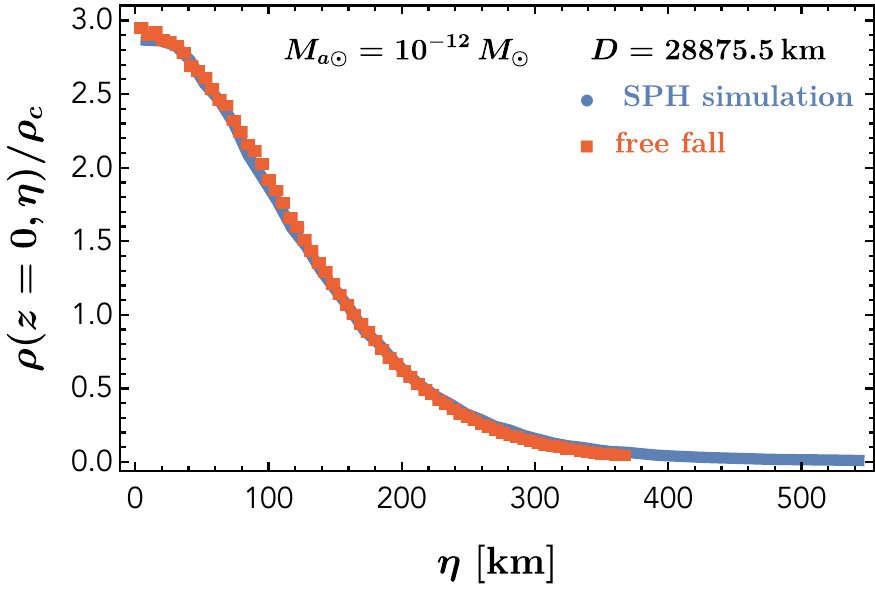}  
    \hfil\hfil
 \end{center}
 \caption{The comparison between the SPH simulation (blue) and free-fall particle approximation (red) at snapshot 3 in Fig.~\ref{fig:slice_M12} ($D=28875.5$km) for $M_{a\odot}=10^{-12} \,M_{\odot}$ and $m_a=10^{-5}$ eV. The vertical axis is normalized by $\rho_c=3\times10^{-22}M_\odot/\text{km}^3$. 
  }
  \label{fig:comparison_M12}
 \end{figure}

Given initial conditions above, we perform the semi-analytic calculation based on the free-fall particle approximations. 
Details about the semi-analytic computations can be found in Appendix \ref{sec:approximation}.
In Figs.~\ref{fig:comparison} and \ref{fig:comparison_M12}, we compare the free-fall particle approximations with the numerical simulation for $M_{a\odot}=10^{-9} \,M_{\odot}$ and $10^{-12} \,M_{\odot}$, respectively.
We observe that the error of free-fall particle approximation is about $30$\% for $M_{a\odot}=10^{-9} \,M_{\odot}$ while less than $10$\% for $M_{a\odot}=10^{-12} \,M_{\odot}$.
It is expected that the error of the free-fall approximation is inversely proportional to the ratio of $T_{\rm cross}/T_{\rm dyn}$.
Therefore, according to Fig.~\ref{fig:timeratio}, the free-fall particle approximation is better for more diluted axion stars.
This expectation is consistent with the results in Figs.~\ref{fig:comparison} and \ref{fig:comparison_M12}.
We conclude that the free-fall particle approximation is trustable for the phenomenologically interesting region, $M_{a\odot}=10^{-15} \,M_{\odot}$.

 \begin{figure}[th!]
 \begin{center}
    \includegraphics[width=.48\textwidth]{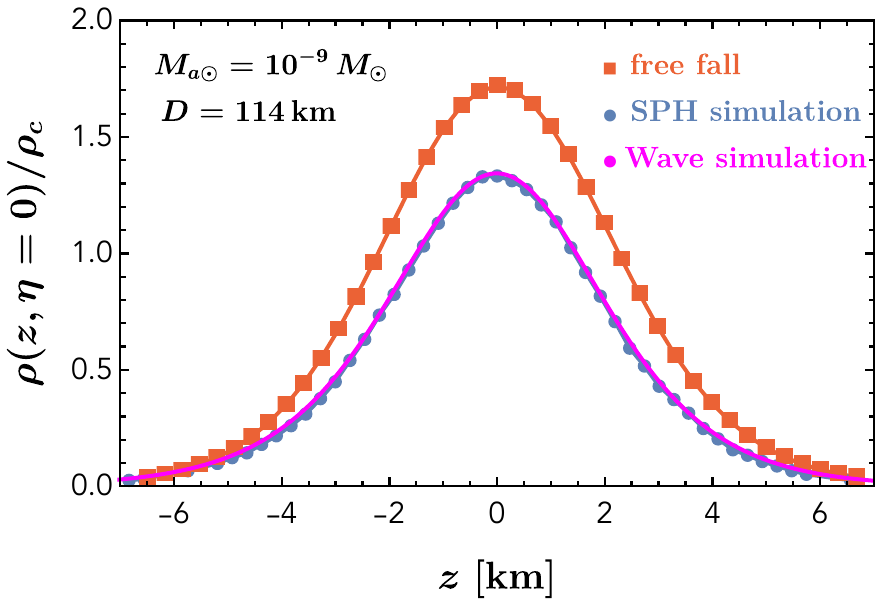} \hspace{3mm}
    \includegraphics[width=.48\textwidth]{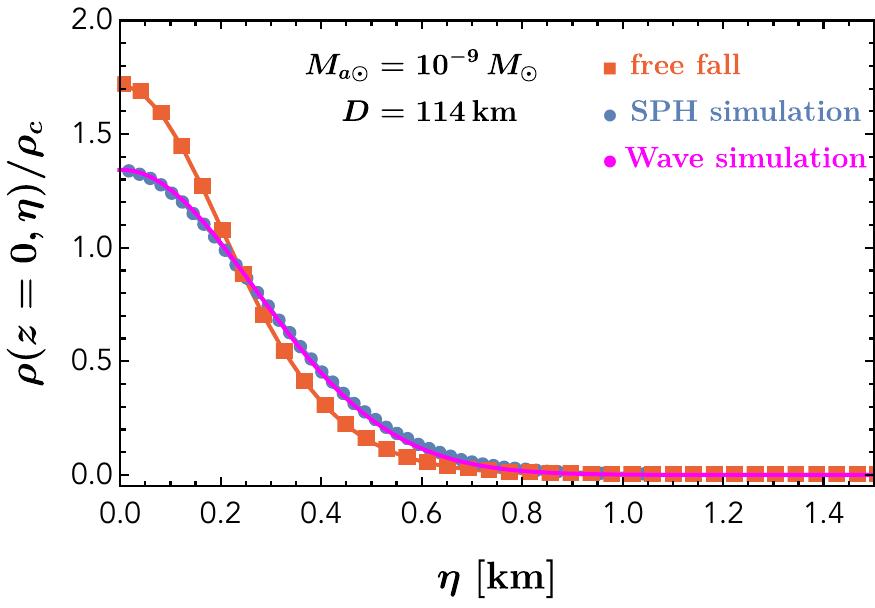}  
    \hfil\hfil
 \end{center}
 \caption{The comparison of the energy density among the wave (magenta), SPH (blue) simulations and semi-analytic computations under the free-fall particle approximation (red) at snapshot 3 in Fig.~\ref{fig:slice} ($D=114$\,km) for $M_{a\odot}=10^{-9} \,M_{\odot}$. The axion mass is taken to be $m_a=10^{-5}$ eV. The vertical axis is normalized by $\rho_c=3\times10^{-10}M_\odot/\text{km}^3$.
  }
  \label{fig:comparison}
 \end{figure}
 
\begin{figure}[th!]
 \begin{center}
 \includegraphics[width=.9\textwidth]{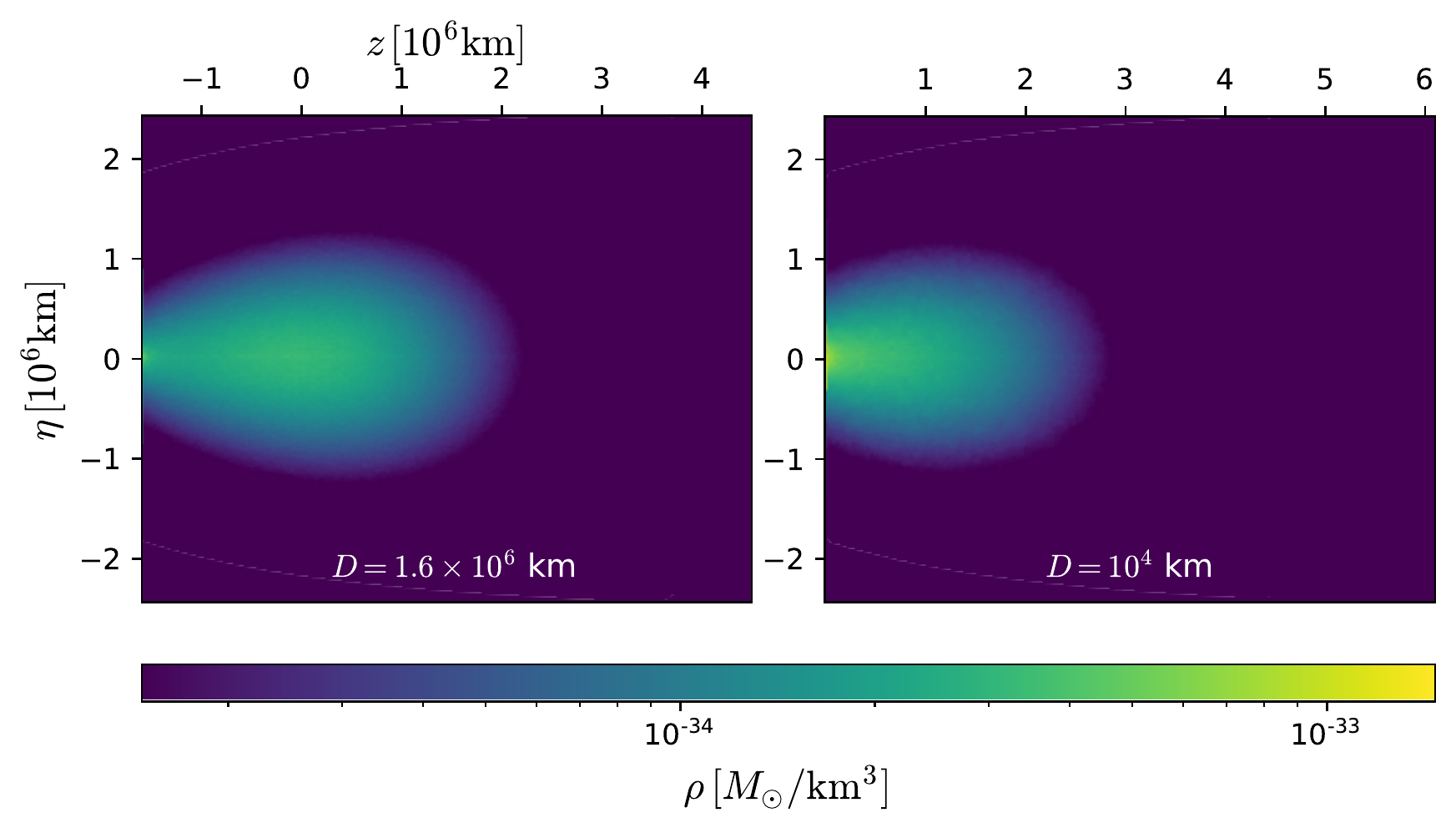}
    \hfil\hfil
\end{center}
\caption{Head-on collision of an axion star of mass $M_{a\odot}=10^{-15} \,M_{\odot}$ with $m_a = 10^{-5}$\,eV and a neutron star of mass $1\,M_{\odot}$ from the free-fall particle approximation calculations. Left panel: slice density when the front end of the axion star reaches the neutron star. Right panel: slice density when the center of axion star reaches the neutron star. The axion star shrinks in $\eta$ direction when it is close to the neutron star, but its size along $\eta$ is still much larger than the radio transition radius (cf. Figs. \ref{fig:slice} and \ref{fig:slice_M12}).
}
\label{fig:density_M15}
\end{figure}

In Fig.~\ref{fig:density_M15}, we show the slice density for $M_{a\odot}=10^{-15} M_{\odot}$. 
In this case, only a small fraction of the axion particles reach the $100$ km region where the resonant conversion takes place. In the next subsection, we compute the fraction within the free-fall particle approximation.
 
\subsection{Fraction of axion particles having resonance conversion}

\begin{figure}[t!]
 \begin{center}
    \includegraphics[width=.47\textwidth]{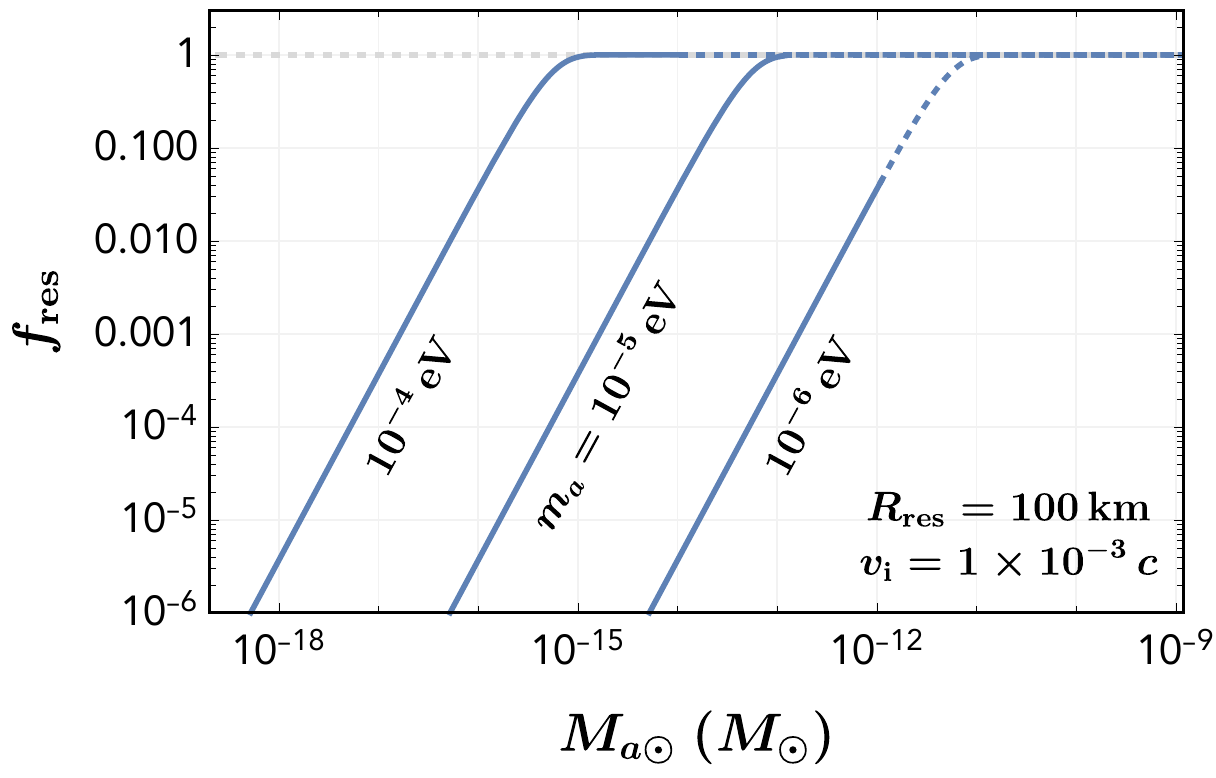}  \hspace{3mm}
      \includegraphics[width=.46\textwidth]{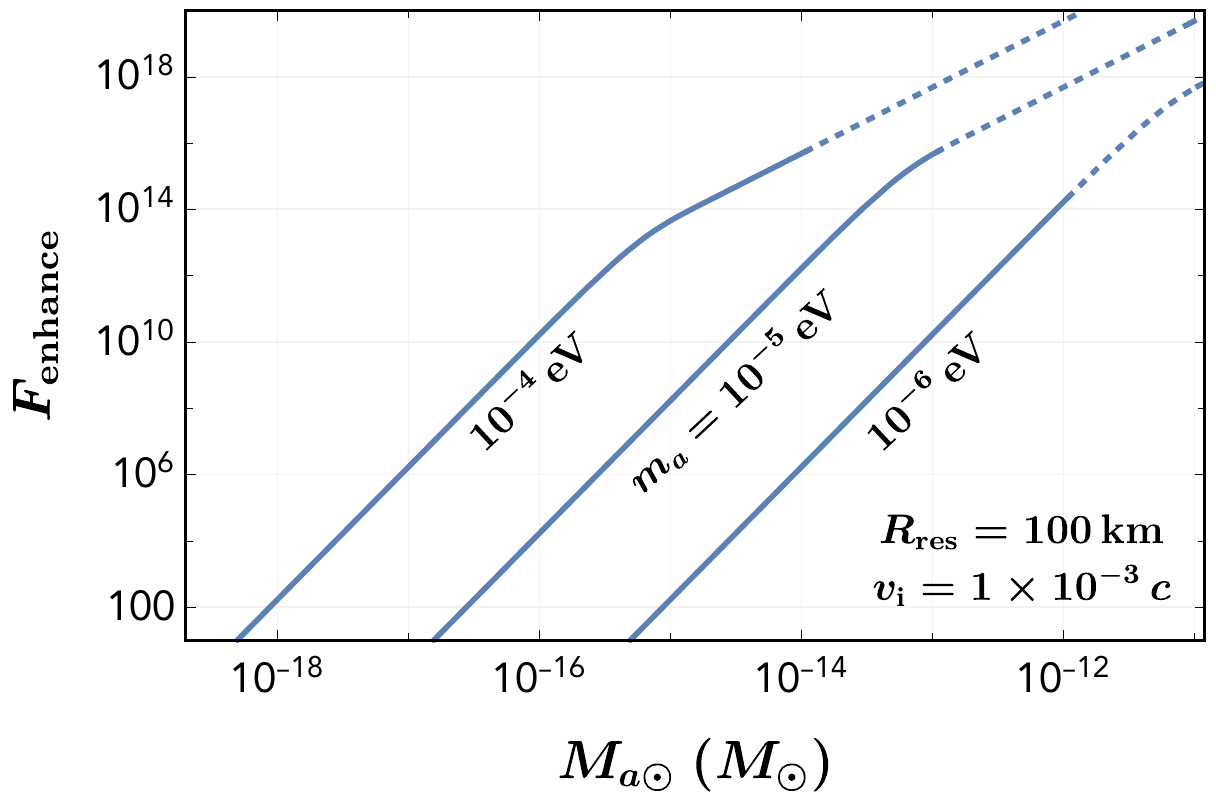}
     \hfil\hfil
 \end{center}
 \caption{Left panel: the fraction of axion particles in an axion star to reach the plasma resonant conversion radius of $R_{\rm res} = 100$~km. The solid line regions are the diluted axion branches to satisfy Eq.~\eqref{eq:diluted-mass-condition}. Right panel: the enhancement factor compared to the signal strength from the averaged axion particles. }
  \label{fig:fraction}
 \end{figure}

Using the analytic formula for a free-fall particle in the neutron star gravitational potential and in the limit of $\eta\ll G_N\,M_{\rm NS}/v_{\rm i}^2$, the minimum distance to the center of the neutron star of a particle with $\eta$ and an initial velocity $v_{\rm i}$ in the $-z$ direction is 
\beqa
R_{\rm min} = \frac{\eta^2\,v_{\rm i}^2}{2\,G_N\,M_{\rm NS}} ~. 
\eeqa
Requiring $R_{\rm min} \le R_{\rm res}$ provides an upper bound on $\eta$
\beqa
\eta < \eta_{\rm max} = \frac{\sqrt{2\,G_N\,M_{\rm NS}\,R_{\rm res}}}{v_{\rm i}} ~.
\eeqa
The fraction of axion particles integrating from $0\le \eta \le \eta_{\rm max}$ is 
\beqa
f_{\rm res}  = 1 - \frac{1}{\left(1+\alpha^2 \,\eta_{\rm max}^2/R_c^2\right)^{13/2}}  \approx  \frac{13 \,\alpha^2 \eta_{\rm max}^2}{2\,R_c^2}~,
\label{eq:f_res}
\eeqa
where the limit of $\alpha\,\eta_{\rm max} \ll R_{\rm AS}$ has been taken for the approximation. Here, $\alpha \approx 0.3$ and $R_c \approx R_{\rm AS}/3.9$.  Therefore, for a diluted axion star with $T_{\rm cross} \ll T_{\rm dyn}$, the fraction of axion particles to reach the plasma resonant conversion radius is well approximated by (we fix $M_{\rm NS}=M_\odot$) 
\beqa
f_{\rm res} &\approx& \frac{0.16\,G_N^3\,M_{\rm NS}\,m_a^4\,M^2_{a\odot}\,R_{\rm res}}{v_{\rm i}^2} \nonumber \\
&=& \left(3.5\times 10^{-4}\right) \left( \frac{m_a}{10^{-5}\,\mbox{eV}} \right)^4 \, \left( \frac{M_{a\odot}}{10^{-15}\,M_\odot} \right)^2\,\left( \frac{R_{\rm res}}{100\,\mbox{km}} \right)\, \left( \frac{1\times 10^{-3}}{v_{\rm i}} \right)^2 ~.
\eeqa
We show the numerical values of the fraction in the left panel of Fig.~\ref{fig:fraction} as a function of axion star masses for different $m_a$. For a denser axion star with a smaller $R_{\rm AS}$, 100\% of axions can reach the resonance-conversion region. 

Noting that we have only considered the head-on collision so far. For a non-zero impact parameter $b$,  the fraction is reduced and becomes
\beqa
f_{\rm res} &=& \frac{13}{2\pi} \int_0^{\tilde{\eta}_{\rm max}} d\tilde{\eta} \int^{2\pi}_0 d\theta \frac{\tilde{\eta}}{\left(1+\tilde{b}^2+\tilde{\eta}^2-2\,\tilde{b}\,\tilde{\eta}\,\sin\theta\right)^{15/2}} ~,
\eeqa
with $\tilde{b}=\alpha\, b/R_c$ and $\tilde{\eta}_{\rm max}=\alpha\,\eta_{\rm max}/R_c$. In the limit of $b \gg R_c$ or $\tilde{b} \gg 1$, the fraction scales like $b^{-15}$ and drops very quickly. 

Before we discuss the converted radio signals, we also estimate the passing time of axion stars off a neutron star, which is given by
\beqa
t_{\rm pass} \approx \frac{2 \,R_{\rm AS} }{ v_{\rm i} } \approx (1.8\times 10^4\,{\rm s})  \left( \frac{10^{-5}\,\mbox{eV}}{m_a} \right)^2 \,  \left( \frac{10^{-15}\,M_\odot}{M_{a\odot}} \right)\, \left( \frac{1\times 10^{-3}}{v_{\rm i}} \right) ~.
\eeqa
So, for a very diluted axion star, the converted radio signal (if above the telescope sensitivity) could last hours or days.  For one axion star and neutron star encounter event, the rate of entering the resonance conversion region is 
\beqa
\Gamma_{a\odot} \approx  \dfrac{f_{\rm res}\,M_{a\odot}}{m_a\,t_{\rm pass}} \approx (2.4\times 10^{48}\,\mbox{s}^{-1})\,\left(\frac{m_a}{10^{-5}\,\mbox{eV}} \right)^5 \,  \left( \frac{M_{a\odot}} {10^{-15}\,M_\odot}\right)^4\,\left( \frac{R_{\rm res}}{100\,\mbox{km}} \right)\, \left( \frac{1\times 10^{-3}}{v_{\rm i}} \right) ~.
\eeqa

One can compare this rate for the axion star case with the averaged axion particle case, and define the enhancement factor as
\beqa
F_{\rm enhance} \approx \dfrac{\Gamma_{a\odot}}{\Gamma_{\rm average}} \,=\, \dfrac{\dfrac{f_{\rm res}\,M_{a\odot}}{m_a\,t_{\rm pass}}}{\dfrac{\rho_{a}}{m_a}\,\pi\,R_{\rm res}^2\,v_{\rm i}\,\left(1+ \dfrac{2\,G_N\,M_{\rm NS}}{v_{\rm i}^2\,R_{\rm res}}\right)} ~,
\eeqa
after taking into account of the ``gravitation focus" effect or the increasing Safronov number~\cite{1972epcf.book.....S}. Using $\rho_a \approx \rho_{\rm DM} \approx 0.4\,\mbox{GeV}/\mbox{cm}^3$ for the local averaged axion energy density, we show the enhancement factor as a function of the axion star mass in the right panel of Fig.~\ref{fig:fraction}. 
 
\section{Searching for axion stars in radio telescopes}
\label{sec:pheno}
In this section, we investigate the possibility to detect the radio wave signal by radio telescopes such as the Green Bank Telescope and the Square Kilometer Array. 

\subsection{Encounter rates}
\label{sec:encounter}

Although the signal strength for one axion star and neutron star encounter event can be large, the frequency for such encounter event with one specific neutron star is too small. Therefore, one could search for converted radio signals for axion stars encountering a collection of neutron stars in our galaxy or nearby galaxies. 

Using the geometric encounter cross section $\sigma = \pi (R_{\rm AS} + R_{\rm res})^2 \approx R_{\rm AS}^2$ for a diluted axion star and the gravitational focus factor, the the encounter rate is
\beqa
\Gamma_{\rm encounter}&\approx& \dfrac{f_{a\odot}\,\rho_{\rm DM}}{M_{a\odot}}\,\times f_{\rm NS}\,N_{\rm NS} \times \sigma v_{\rm i} \, \times \, \left(1+ \dfrac{2\,G_N\,M_{\rm NS}}{v_{\rm i}^2\,R_{\rm AS}}\right) \\
 && \hspace{-2.5cm}\approx
\begin{cases}
 (167\,\mbox{day}^{-1}) \, f_{a\odot}\,f_{\rm NS}\,\left( \dfrac{10^{-5}\,\mbox{eV}}{m_a} \right)^4 \,  \left( \dfrac{10^{-15}\,M_\odot}{M_{a\odot}} \right)^3\, \left( \dfrac{v_{\rm i}}{1\times 10^{-3}} \right)   & ~\mbox{for} \quad R_{\rm AS} \gg \dfrac{2\,G_N\,M_{\rm NS}}{v_i^2}\\
 (192\,\mbox{day}^{-1}) \, f_{a\odot}\,f_{\rm NS}\,\left( \dfrac{10^{-5}\,\mbox{eV}}{m_a} \right)^2 \,  \left( \dfrac{10^{-15}\,M_\odot}{M_{a\odot}} \right)^2\, \left( \dfrac{1\times 10^{-3}}{v_{\rm i}} \right)   & ~\mbox{for}\quad R_{\rm AS} \ll \dfrac{2\, G_N\,M_{\rm NS}}{v_i^2}
 \end{cases} ~, \nonumber 
\eeqa
where we have used $\rho_{\rm DM} \approx 0.3\,\mbox{GeV}/\mbox{cm}^3$ and $N_{\rm NS} \simeq 10^{9}$ as the rough numbers for the Milky Way galaxy and introduced $f_{a\odot}$ as the fraction of axion stars in the total axion energy density and $f_{\rm NS}$ as the fraction of neutron stars with a large enough magnetic field and close distance to the telescope to be detected by one telescope.

To have a more precise estimation of the event rates, we take into account of the dark matter (axion star) and neutron star distributions. Two dark matter profiles will be considered for the Milky Way (MW) and Andromeda (M31) galaxies. One is the NFW profile given by~\cite{Navarro:1996gj} 
\beqa
\rho_{\rm NFW}(r) = \frac{\rho_0}{\frac{r}{R_s} \left( 1 + \frac{r}{R_s} \right)^2} ~, 
\eeqa
with $R_s = 20 (16.5)$~kpc for MW(M31).  The normalization factor $\rho_0$ for MW is determined by the local dark matter density $\rho(R_\odot = 8.5\,\mbox{kpc}) = 0.3\,\mbox{GeV}/\mbox{cm}^3$. For M31, one has $\rho_0 = 0.418\,\mbox{GeV}/\mbox{cm}^3$~\cite{Tamm_2012}. The other profile we consider is the cored Burkert profile~\cite{Burkert:1995yz} 
\beqa
\rho_{\rm core} (r) = \frac{\rho_0}{\left( 1 + \frac{r}{R_s} \right) \left[ 1 +\left( \frac{r}{R_s}\right)^2 \right]} ~,
\eeqa
with the core radius $R_s = 9$~kpc for both MW and M31~\cite{Tamm_2012} and $\rho_0 = 1.77\,\mbox{GeV}/\mbox{cm}^3$ for the M31.

We also consider the globular cluster M54 at the center of the nearby Sagittarius dwarf galaxy assuming a cuspy NFW profile or a cored profile.
The NFW profile has the scale radius $R_s \approx 0.2$~kpc with the normalization factor  $\rho_0\,R_s^3 \approx 3.3\times 10^7 \,M_\odot$~\cite{HESS:2007ora}. For the cored profile, the isothermal one will be used and has 
\beqa
\rho_{\rm core} = \frac{v_a^2}{4\pi\,G_N}\, \frac{3 R_s^2 + r^2}{(R_s^2 + r^2)^2} ~,
\eeqa
with $R_s \approx  230$~pc (see Ref.~\cite{Safdi:2018oeu} for related discussion) and $v_a\approx 13.4\,\mbox{km}/\mbox{s}$.

For the neutron star distribution in the MW and M31, the galaxy bulge component has~\cite{Edwards:2020afl}
\beqa
n_{\rm bulge}(\eta, z) = N_{\rm bulge}\, \frac{11.1}{\mbox{kpc}^3}\, \frac{e^{- (\eta^2 + z^2/q^2)/r_{\rm cut}^2}}{ \left(1 + \sqrt{\eta^2 + z^2/q^2}/r_0 \right)^\lambda} ~. 
\eeqa
with $q=0.5$, $\lambda = 1.8$, $r_0 = 0.075$~kpc and  $r_{\rm cut} = 2.1$~kpc. The distribution in the disk follows the Lorimer profile and has~\cite{Lorimer:2006qs,Edwards:2020afl}
\beqa
n_{\rm disk}(\eta, z) = N_{\rm disk} \, \frac{C^{B+2} \, e^{-C} }{4\pi\,R_\odot^2\,\sigma_z\,\Gamma(B+2)}\,\left( \frac{\eta}{R_\odot}\right)^B \, e^{- C\, \frac{\eta - R_\odot}{R_\odot}}\,e^{ - \frac{|z|}{\sigma_z}} ~,
\eeqa
with $B=3.91$, $C=7.54$ and $\sigma_z = 0.76$~kpc. We take $N_{\rm bulge}=4.8\times 10^8$ and $N_{\rm disk}=3.2\times 10^8$ for the MW and $N_{\rm bulge}=2\times 10^9$ and $N_{\rm disk}=4\times 10^8$ for the M31. 

For M54 and following Ref.~\cite{Safdi:2018oeu}, we take totally around 2400 neutron stars with 1152 in the inner core with a constant density up to $a \sim 0.7$~pc and 1248 neutron stars following the Plummer sphere model for $r > a$. 

\begin{figure}[t!]
 \begin{center}
    \includegraphics[width=.47\textwidth]{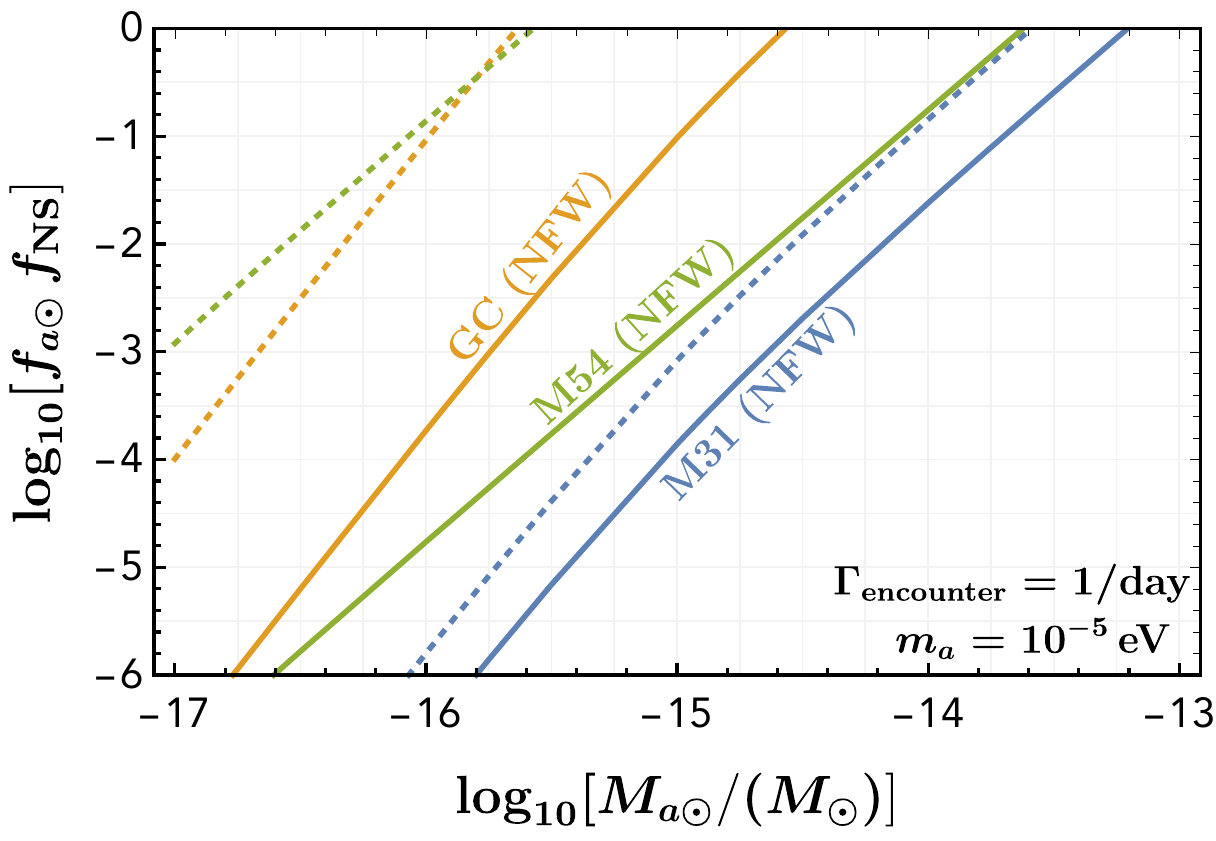} \hspace{0.4cm}
        \includegraphics[width=.47\textwidth]{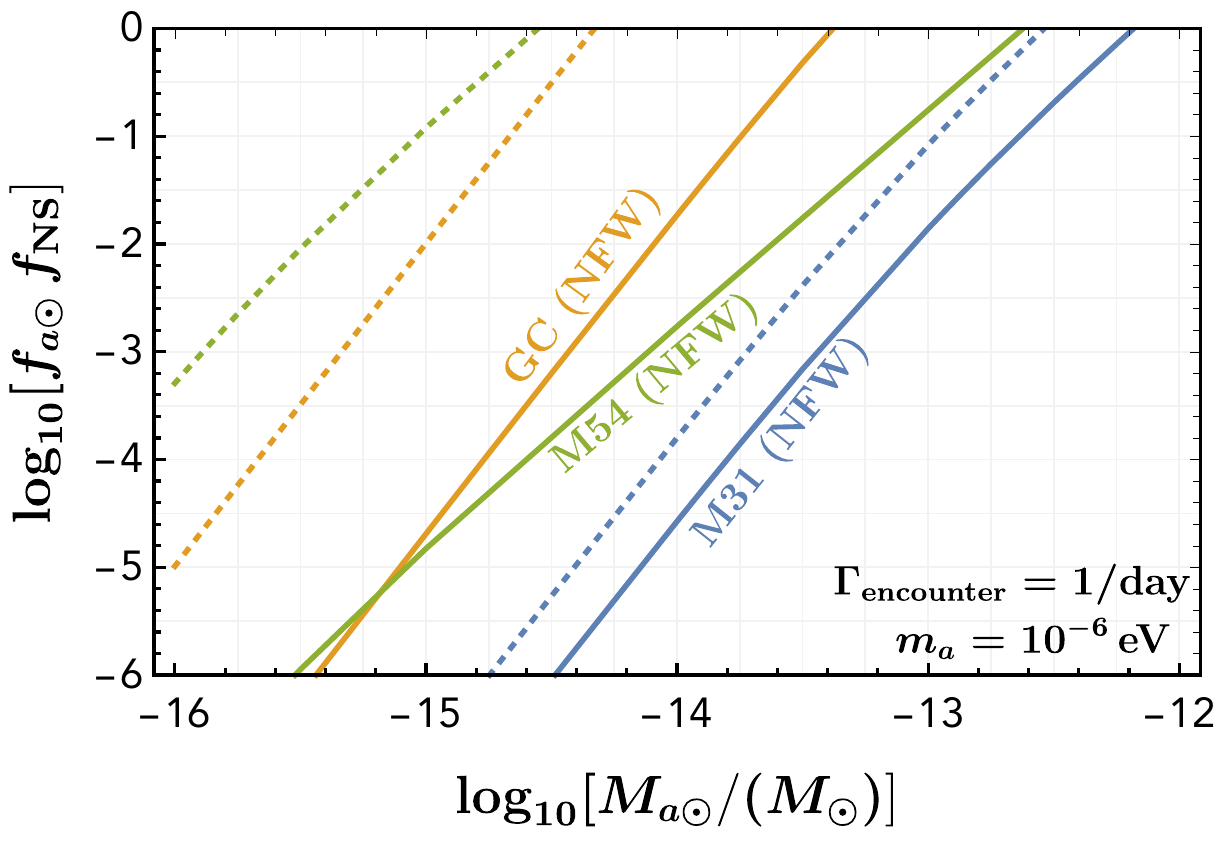}  
     \hfil\hfil
 \end{center}
 \caption{The parameter region to have one encounter event per day. The solid lines are for the dark matter NFW profiles, while the dotted lines are for the dark matter cored profiles. For the MW galactic center (GC), an observation angular radius $0.1^\circ$ has been imposed.}
  \label{fig:encounter}
 \end{figure}

The encounter rate follows an integration 
\beqa
\Gamma_{\rm encounter}&\approx&f_{a\odot}\, f_{\rm NS}\, \int dV \dfrac{\rho_{\rm DM}}{M_{a\odot}}\,\times \,n_{\rm NS}\times R_{\rm AS}^2\, v_{\rm i} \, \times \, \left(1+ \dfrac{2\,G_N\,M_{\rm NS}}{v_{\rm i}^2\,R_{\rm AS}}\right)  ~.
\eeqa
Here, we fix $v_{\rm i} = 10^{-3}\,c$ for the MW and M31 and $v_{\rm i} = v_a\approx 13.4\,\mbox{km}/\mbox{s}$ for the M54. By requiring one encounter event for one-day telescope observation, in Fig.~\ref{fig:encounter} we show the fraction product $f_{a\odot} f_{\rm NS}$ as a function of the axion star mass for $m_a = 10^{-5}$\,eV in the left panel and $m_a = 10^{-6}$\,eV in the right panel. 

 %

\subsection{Radio signals}
\label{sec:radio}

In the presence of the plasma, the photon acquires a mass corresponding to the plasma frequency. 
Within the Goldreich and Julian model~\cite{Goldreich:1969sb}, the resonance conversion radius $R_{\rm res}$ where the plasma frequency equals to the axion mass is~\cite{Hook:2018iia,Huang:2018lxq}
\beqa
R_{\rm res}(\theta,\theta_m,t)=224\,\mbox{km} \times|3\cos\theta\,\hat{m}\cdot\hat{r}-\cos\theta_m|^{1/3} \left(\frac{R_{\rm NS}}{10\,\mbox{km}}\right) \left[\left(\frac{B_0}{10^{14}\text{G}}\right)\left(\frac{1\,\text{s}}{P}\right)\left(\frac{\text{1\,GHz}}{m_a}\right)^2\right]^{1/3}  ~.
\eeqa
Here, the axion mass corresponding radio frequency is $\nu = m_a/(2\pi) = 2.4\,\mbox{GHz}\times (m_a/10^{-5}\,\mbox{eV})$; $\theta_m$ is the angle between the magnetic dipole axis and the rotation axis; $B_0$ is the magnetic field at the pole of the neutron star with the outside dipole magnetic field $|\vec{B}(\vec{r})|=B_0(R_{\rm NS}/r)^3$; $P$ is the neutron star spin period. Both $B_0$ and $P$ follow the log-normal distributions. For $B_0$ in the unit of Gauss, it has 
\beqa
\label{eq:B-distribution}
p(B_0) = \dfrac{1}{\sqrt{2\pi}\,\sigma_{\log_{10}B_0}} \, \mbox{exp}\left[ - \dfrac{\left(\log_{10} B_0 - \langle \log_{10} B_0 \rangle\right)^2}{2\,\sigma^2_{\log_{10}B_0}} \right] ~,
\eeqa
with $\langle \log_{10} B_0 \rangle = 12.65$ and $\sigma_{\log_{10}B_0} = 0.55$~\cite{FaucherGiguere:2005ny,Bates:2013uma}. For $P$, it has $\langle \log_{10} P/\mbox{ms} \rangle = 2.7$ and $\sigma_{\log_{10}P/\mbox{ms}} = 0.34$~\cite{Lorimer:2006qs}. To simplify our calculations, we will choose $\theta_m =0$ and averaging the $\theta$-dependent part to have $\langle |3\cos\theta\,\hat{m}\cdot\hat{r}-\cos\theta_m|^{1/3} \rangle \approx 1.7$. For $m_a \lesssim 5 \times 10^{-3}\,\mbox{eV}$, $R_{\rm res} > 10$\,km and is outside the neutron star radius.

Based on the WKB approximation, the resonant conversion probability from axion to photon has the following formula~\cite{Raffelt:1987im,Hook:2018iia,Millar:2021gzs}
\beqa
p_{a\gamma}^{\infty} \approx g_{a\gamma\gamma}^2 \, B^2(R_{\rm res})\, \frac{\pi\,R_{\rm res}}{3\,m_a} ~.
\eeqa 

By taking the time average of the converted photon energy, we define the averaged radiation power as 
\beqa
\label{eq:power}
\left\langle\frac{d\mathcal{P}}{d\Omega}\right\rangle= 2\,p_{a\gamma}^\infty \,\frac{f_{\rm res}\,M_{a\odot}}{t_{\rm pass}} ~.
\eeqa
The time-averaged spectral density flux $\langle S \rangle$ is
\beqa
\langle S \rangle &=& \frac{1}{\text{BW}} \frac{1}{4\pi d_{\rm source}^2} \left\langle\frac{d\mathcal{P}}{d\Omega}\right\rangle = (13\,\mu\mbox{Jy})\, \left(\frac{g_{a\gamma\gamma}}{10^{-12}\,\mbox{GeV}^{-1}} \right)^2 \left(\frac{m_a}{10^{-5}\,\mbox{eV}} \right)^5 \,  \left( \frac{M_{a\odot}} {10^{-15}\,M_\odot}\right)^4 \nonumber \\
&&\hspace{3.5cm}\times  \left( \frac{B(R_{\rm res})}{10^{10}\,\mbox{G}} \right)^2 \left( \frac{R_{\rm res}}{100\,\mbox{km}} \right)^2\, \left( \frac{1\times 10^{-3}}{v_{\rm i}} \right)\,\left( \frac{1\,\mbox{kHz}}{\text{BW}} \right)\,\left( \frac{10\,\text{kpc}}{d_{\rm source}} \right)^2 ~.
\eeqa
Here, $d_{\rm source}$ is the distance from the telescope to the neutron star. We take $d_{\rm source}$ to be 8.5~kpc for the MW galactic center (GC), 780~kpc for the M31 and 25~kpc for the M54. 

For the signal from an individual neutron star, we will simply choose the bandwidth to be $\text{BW}=1$\,kHz, which is the resolution of the current and planned ratio telescopes~\cite{Perley:2011ur,Braun:2019gdo}. There are still uncertainties to understand the proper choice of BW. For instance, Ref.~\cite{Battye:2021xvt} has pointed out that the Doppler shift effects dominate the signal frequency broadening over the one from dark matter velocity dispersion. For the diluted axion star considered here, the axion particles in the initial axion star state are in a coherent BEC state. Their velocity dispersion is much smaller than the free axion particle case. One could apply the ray-tracing method to the axion star case and obtain a more precise value for BW.

The sensitivity on the spectral flux density for different telescopes has~\cite{Safdi:2018oeu,Battye:2019aco}
\beqa
S_{\rm min} = \text{SNR}_{\rm min} \, \frac{\sqrt{2}\,T_{\rm sys} }{\sqrt{\text{BW}}\,\sqrt{t_{\rm obs}}\,A_{\rm eff}} ~, 
\eeqa
Here, $\text{SNR}_{\rm min}$ is the minimal signal-to-noise ratio; $A_{\rm eff}$ is the effective area of the telescope; $t_{\rm obs}$ is the telescope observation time for a certain astrophysical region and taken to be $t_{\rm obs}=1$\,day in our later analysis; the temperature $T_{\rm sys}$ is the summation of the telescope-specific one $T_R$ and the astrophysical one  $T_{\rm astro}$. For the astrophysical one, we take $T_{\rm astro}=10$\,K for the M31 and M54, and $T_{\rm astro}=10^3$\,K for the MW galactic center. For the GBT, we take $A_{\rm eff} \approx 5500\,\mbox{m}^2$ and $T_R=25$\,K up to $\nu \approx 40$~GHz~\cite{GBT_TR}. For the SKA2 with the $A_{\rm eff}$ up to $\sim 10^6\,\mbox{m}^2$, we use the values of $A_{\rm eff}/T_R$ in Ref.~\cite{SKAinfo}. Note that the SKA2 can cover the frequency range from 0.05\,GHz to 40\,GHz, which is translated into the axion mass from $2\times 10^{-7}$\,eV to $1.7\times 10^{-4}$\,eV. 
Numerically, the value of $S_{\rm min}$ is estimated to be
\beqa
S_{\rm min}  
\approx \left(10^4 \, \mu\mbox{Jy}\right)\,\left(\frac{\text{SNR}_{\rm min}}{5}\right)\,\left(\frac{100\,\mbox{m}^2/\mbox{K} }{A_{\rm eff}/T_{\rm sys}}\right) \,\left(\frac{\mbox{kHz}}{\text{BW}}\right)^{1/2} \,\left(\frac{\mbox{day}}{t_{\rm obs}}\right)^{1/2} \,.
\eeqa

\begin{figure}[t!]
 \begin{center}
    \includegraphics[width=.475\textwidth]{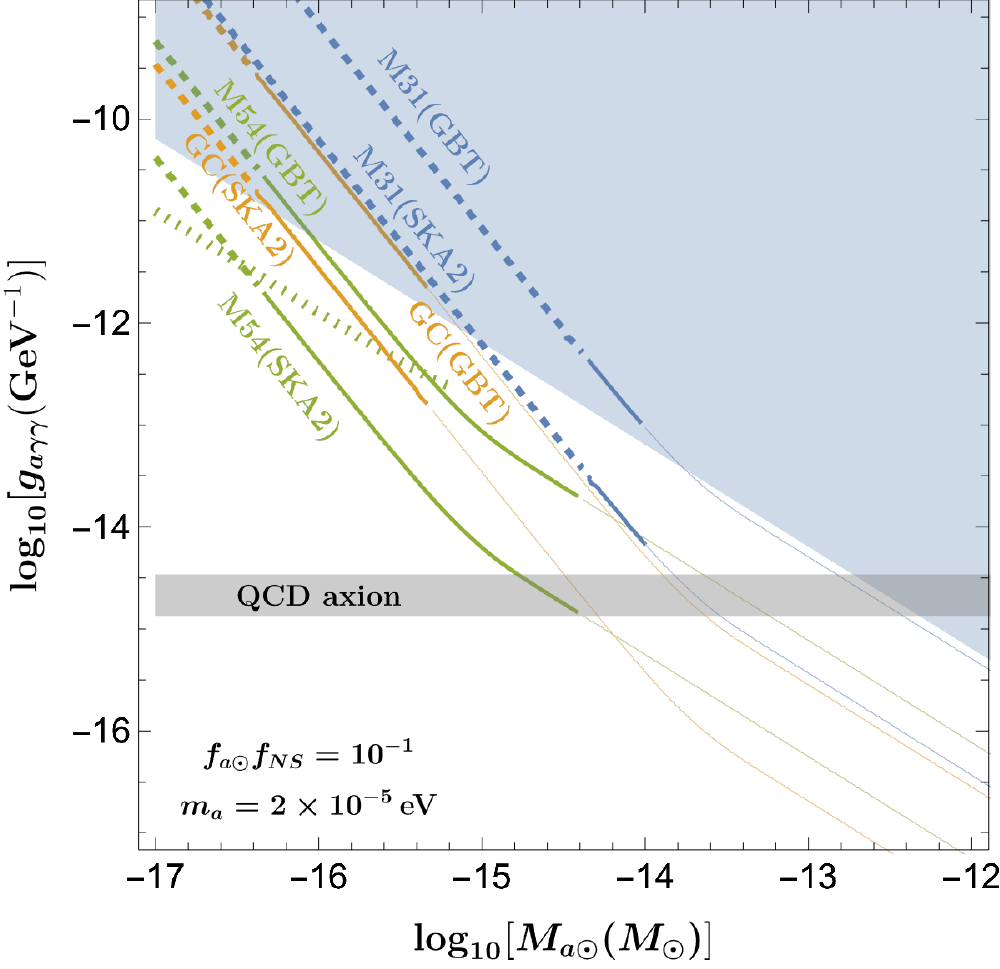}  \hspace{3mm}
      \includegraphics[width=.48\textwidth]{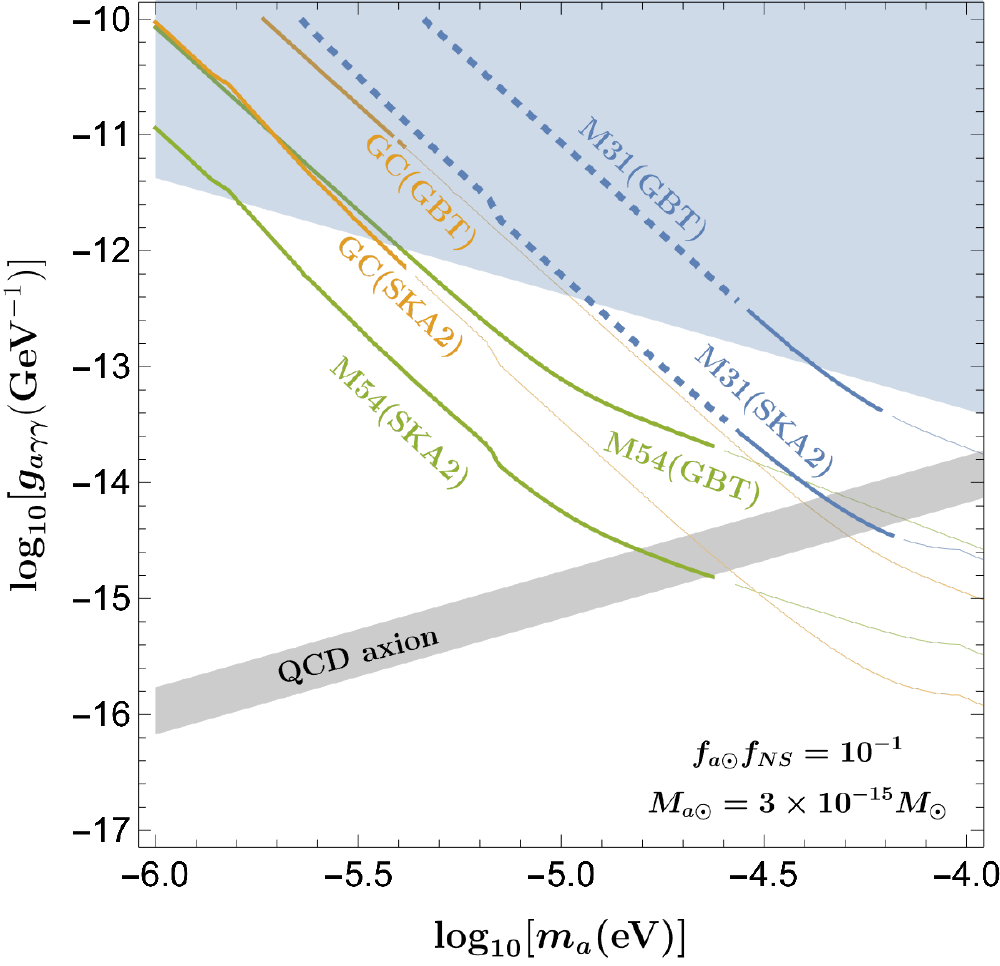} 
     \hfil\hfil
 \end{center}
 \caption{The axion mass and coupling parameter space that can be probed by current and future radio telescopes. The blue shaded region does not satisfy the condition in Eq. \eqref{eq:diluted-mass-condition} to be in the diluted axion star branch. The gray shaded region is the QCD axion representative region. Different lines correspond to $\langle S\rangle/S_{\rm min}=1$ with the regions above the lines are constrained by different telescope data. Different dark matter profiles have different encounter rates and upper axion star masses according to Fig.~\ref{fig:encounter}. The dashed (solid) line is for the cored (NFW) profile. The dotted line corresponds to constraints from an ensemble of encounter events. The band width is fixed to be $\text{BW}=1$\,kHz. Left panel: $(M_{a\odot}, g_{a\gamma\gamma})$-plane. Right panel: $(m_a, g_{a\gamma\gamma})$-plane. Both panels have a fixed $f_{a\odot}f_{\rm NS} = 10^{-1}$ and $t_{\rm obs}=1$\,day. 
}
  \label{fig:detectability}
 \end{figure}

Requiring $S \geq S_{\rm min}$, in the left panel of Fig.~\ref{fig:detectability} we show the GBT and SKA reaches in the parameter space of $M_{a\odot}$ and $g_{a\gamma\gamma}$ with $m_a= 2\times 10^{-5}$\,eV from observing M31, M54 and MW GC for one day. Here, we fix $B_0=10^{13.2}\,\text{G}$ ($P=10^{0.04}\,\text{s}$), which are one sigma above (below) the averaged value in Eq.~\eqref{eq:B-distribution}. The telescope reaches are not sensitive to this choice. We also fix $\text{BW}=1$\,kHz, $\text{SNR}_{\rm min}=5$, $t_{\rm obs}=1$\,day and $f_{a\odot}f_{\rm NS}=0.1$. For a smaller value of $f_{a\odot}f_{\rm NS}$, the end points of the dashed (for the cored dark matter profile) and solid (for the NFW profile) lines shift in the smaller $M_{a\odot}$ direction. Increasing the observational time $t_{\rm obs}$, on the other hand, one shifts the lines downward and pushes the end points rightward. One can see that for the SKA2 telescope, the search for the encounter events in the M54 could probe the QCD axion parameter region for the axion star mass slightly above $10^{-15}\,M_{\odot}$. In the right panel, we fix $M_{a\odot} = 3\times 10^{-15}\,M_\odot$ and show the telescope reaches in terms of $m_a$ and $g_{a\gamma\gamma}$ by choosing the same values for other parameters. One can see that for a heavy axion mass close to $10^{-4}$\,eV and the NFW dark matter profile, the observation of M31 by the SKA2 could also probe the QCD axion model parameter space.

\section{Discussion and conclusions}
\label{sec:conclusion}

We have studied the tidal evolution of axion stars during its collision with a neutron star using both numerical simulations and free-fall particle approximation. For different mass ranges of axion stars with a fixed axion particle mass $m_a = 10^{-5}$\,eV, we have employed different numerical and analytic methods: pseudo-spectral simulations for a relatively compact axion star, $M_{a\odot}\gtrsim10^{-9} M_{\odot}$, SPH simulations for axion stars with a mass between $10^{-9} M_{\odot}$ and $10^{-12} M_{\odot}$, and the free-fall particle approximation for axion stars with a mass smaller than $10^{-12} M_{\odot}$. Interestingly, we find that in the case where the internal dynamical time scale of an axion star is much larger than the time it passes the neutron star (see Fig. \ref{fig:timeratio}), e.g. axion star with a mass smaller than $10^{-12} M_{\odot}$, the density evolution of the axion star within the Roche radius can be well approximated by the free-fall assumption, i.e. the axion particles free-fall under the gravitational potential of the neutron star with negligible effects from the self-gravity and quantum pressure. Using the free-fall particle approximation, we calculate the fraction of axion particles in an axion star that can enter the resonance conversion radius (see Eq. \eqref{eq:f_res}).

In the calculation of encounter rate of axion stars with neutron stars, we have assumed that a fraction of dark matter in the Universe, $f_{a\odot}$, is contained in axion stars and taken $f_{a\odot} \sim 1$. Depending on the specified models of how the axion star forms and how they are disrupted by stars and disk in the galaxy \cite{Tinyakov:2015cgg,Dokuchaev:2017psd,Kavanagh:2020gcy,Edwards:2020afl}, $f_{a\odot}$ can be smaller than $1$. In one of the popular models where the Peccei-Quinn symmetry is broken after inflation, the large fluctuations in axion density field collapse in the radiation dominated epoch and form dense clumps called axion miniclusters. It has been shown that axion star can form in the center of axion minicluster through gravitational cooling \cite{Seidel:1993zk,Levkov:2018kau,Eggemeier:2019jsu,Chen:2020cef}. The mass of axion star $M_{a\odot}$ has a power law relation with the virial mass of the axion minicluster $M_{\rm MC}$ \cite{Schive:2014hza}:
\begin{equation}
M_{a\odot} = \left(1.2\times 10^{-16} M_{\odot}\right)\, (1+z)^{1/2}\left(\frac{10^{-5}{\rm eV}}{m_a}\right)\left(\frac{M_{\rm MC}}{10^{-12}M_{\odot}}\right)^{1/3}.
\label{eq:Mstar_halo}
\end{equation}
Note that here we use the definition $M_{a\odot}\sim 4\,M_c \equiv 4\,M(r<R_c)$. Thus only a small fraction of axion particles are contained in the central axion star. Simulations of axion miniclusters \cite{Eggemeier:2019khm} show that at the matter-radiation equality, i.e $z=z_{\rm eq}\sim 3000$, the mass function of axion minicluster $dn/d\ln M \propto M_{\rm MC}^{-0.7}$ and has an exponential cutoff at $M_{\rm MC}=10^{-11} M_{\odot}$. Assuming the relation in Eq.  \eqref{eq:Mstar_halo} and integrating the mass function over the range $(5\times10^{-16}, 10^{-11}) M_{\odot}$, we get $f_{a\odot}\approx0.4\times 0.042 = 0.017$ for $m_a=10^{-5}\,{\rm} {\rm eV}$. Here the lower bound of $M_{\rm MC}$ is the minimum mass of an axion minicluster that can form at a specific redshift, which can be calculated from Eq. \eqref{eq:Mstar_halo} by setting $M_{a\odot}=M_{\rm MC}$. The factor $0.4$ is the fraction of axion particles contained in all axion miniclusters at $z=z_{\rm eq}$ \cite{Eggemeier:2019khm}. At redshift $z \ll z_{\rm eq}$, only a small number of new low-mass axion minicluster forms and the evolution is dominated by mergers of axion miniclusters \cite{Eggemeier:2019khm}. At this stage, small axion miniclusters are accreted onto massive miniclusters and become sub-miniclusters. Some fraction of the axion stars together with the sub-miniclusters will be disrupted by the tidal interaction with the host minicluster or the stars and disk in the galaxy. Thus in the present day $f_{a\odot}$ will be smaller. The exact value is yet to be determined from simulations at lower redshifts, but see Refs.~\cite{Tinyakov:2015cgg,Dokuchaev:2017psd,Kavanagh:2020gcy,Edwards:2020afl} for analytic and semi-analytic treatments. Besides, in some other axion-like particle models, it has been shown that the axion stars can make up a large fraction of dark matter~\cite{Guth:2014hsa,Davidson:2016uok}.

Although Fig.~\ref{fig:detectability} shows that some QCD axion parameter space could be probed by the SKA2 telescope for $M_{a\odot}$ around $10^{-15}\,M_\odot$, there are various large uncertainties including the dark matter profile, the neutron star distributions, the fraction of axion stars in dark matter, the order-of-magnitude estimation of the radiation power in Eq. \eqref{eq:power}, as well as the proper band width adopted by the telescope data analysis. On the other hand, Fig.~\ref{fig:detectability} does demonstrate that the radio telescopes have some chance to discover an axion star that is made of QCD-like axion particles. In our simple-minded telescope reach analysis, we have ignored the possibility of using the time information to relax the requirement of $\mbox{SNR}_{\rm min}$. For instance, one could perform a combined analysis in spectral and time to target on the transient feature of the axion star and neutron star collision event, just like the searches done for the fast radio bursts~\cite{Lorimer:2007qn}. 

\subsection*{Acknowledgments}
The work of Y.B. is supported by the U.S. Department of Energy under the contract DE-SC-0017647.  X.D. acknowledges support from NASA ATP grant 17-ATP17-0120. The work of Y.H. is supported by JSPS Overseas Research Fellowships. Y.H. also thanks the 2021 Simons summer workshop at the Simons Center for Geometry and Physics for the hospitality where part of this work was carried out. X.D. thanks Benedikt Eggemeier for beneficial discussions on the simulation of axion minicluster and kindly providing their simulation results.

\appendix
\section{Pseudospectral method}
\label{sec:ps_method}
Following \cite{Du:2018qor}, we solve the Schr\"{o}dinger-Poisson equations (\ref{eq:schr}) and (\ref{eq:poisson}) using a fourth-order in time pseudospectral method. The wave function is evolved in time by the Hamiltonian operator $H$:
\begin{equation}
\psi(t+\Delta t)=e^{-i H \Delta t}\psi(t) ~.
\label{eq:evo}
\end{equation}
We can split the Hamiltonian into the kinetic operator $K=-\frac{1}{2m_a}\nabla^2$ and the potential operator $W=\Phi_{\rm total}$. To the fourth-order, we have \cite{McLachlan:1995}
\begin{eqnarray}
e^{-i H \Delta t}&\approx&e^{-i v_2 W\Delta t}e^{-i t_2 K\Delta t}e^{-i v_1 W\Delta t}e^{-i t_1 K\Delta t} e^{-i v_0 W\Delta t}\nonumber\\
&&e^{-i t_1 K \Delta t}e^{-i v_1 W \Delta t}e^{-i t_2 K\Delta t}e^{-i v_2 W\Delta t}\,,
\label{eq:exp_H_M}
\end{eqnarray}
where
\begin{eqnarray}
&&v_1=\frac{121}{3924}(12-\sqrt{471}) \,, \quad w=\sqrt{3-12 v_1+9 v_1^2} \,, \quad t_2=\frac{1}{4}\left(1-\sqrt{\frac{9 v_1-4+2 w}{3 v_1}}\right),\nonumber\\
&&t_1=\frac{1}{2}-t_2 \,, \qquad v_2=\frac{1}{6}-4 v_1 t_1^2 \,, \qquad v_0=1-2(v_1+v_2) ~.
\label{eq:M_par}
\end{eqnarray}

In the Cartesian coordinate system, we can make use of the Fourier transform to transform the wave function from the real space to the Fourier space where the kinetic operator is easy to calculate, and transform it back when we need to act the potential operator. Using the fast Fourier transform (FFT) algorithm, this method can be very efficient.

When the system has rotational symmetry, it is easier to work with cylindrical coordinates $(\eta,\varphi,z)$, where $\eta$ is the axial distance, $\varphi$ is the azimuth, and $z$ is the height. The Schr\"{o}dinger-Poisson equations (\ref{eq:schr}) and (\ref{eq:poisson}) then become
\begin{eqnarray}
i\,\frac{\partial \psi(\eta,z)}{\partial t}=-\frac{1}{2m_a}
  \left(\frac{\partial^2}{\partial \eta^2}+\frac{1}{\eta}\frac{\partial}{\partial \eta}+\frac{\partial^2}{\partial z^2}\right)\psi(\eta,z,t)
+ m_a\,\Phi_{\rm total}(\eta,z,t) \,\psi(\eta,z,t) ~,
\label{eq:schr_cy} \\
\left(\frac{\partial^2}{\partial \eta^2}+\frac{1}{\eta}\frac{\partial}{\partial \eta}+\frac{\partial^2}{\partial z^2}\right) \Phi_{\rm self}(\eta,z)=4\pi \, G_N\, m_a\, \rho(\eta,z,t) ~.
\label{eq:poisson_cy}
\end{eqnarray}
Here, $\rho(\eta,z,t)=|\psi(\eta,z,t)|^2$, and we have assumed that the wave function depends only on $\eta$ and $z$. Due to the presence of the term $\frac{1}{\eta}\frac{\partial}{\partial \eta}$, the Laplace operator is no longer diagonal in the Fourier space, so the calculation of $e^{-i K\Delta t}$ is not as easy as in the Cartesian coordinate system. So we first do a Fourier transform only in the $z$ direction
\begin{equation}
X(\eta,z,t) = \sum_{j} X_{k_j}(\eta,t)e^{i k_j z} ~,
\label{eq:X_k}
\end{equation}
where $X=\psi, \Phi, \rho$. Then we have
\begin{equation}
\nabla^2 X(\eta,z,t) = \sum_{j} \left[\left(\frac{\partial^2}{\partial \eta^2}+\frac{1}{\eta}\frac{\partial}{\partial \eta}\right)X_{k_j}(\eta,t)-k^2 X_{k_j}(\eta,t)\right]e^{i k_j z} ~.
\label{eq:X_k_nabla}
\end{equation}
In the $\eta$ direction, we extend the domain from $[0,L_{\eta}/2]$ to $[-L_{\eta}/2,L_{\eta}/2]$ with $L_{\eta}$ a finite number, and discretize the space using Chebyshev-Gauss-Lobatto nodes
\begin{equation}
\eta_i = \frac{L_{\eta}}{2}\cos\left(\frac{i-1}{N_{\eta}}\pi\right)\,, \qquad i=1,\,2,\,...,\,N_{\eta} ~.
\label{eq:eta_i}
\end{equation}
To avoid the coordinate singularity at $\eta=0$, we assume that $\psi$ and $\Phi$ are both even functions and use an even number of grid points to have $\eta_i \ne 0$. To compute the derivative with respect to $\eta$, we make use of the Chebyshev differential matrix~\cite{Bayliss:1994,Baltensperger:1999,Trefethen:2000}
\begin{eqnarray}
D^{\rm cheb}_{11}     &=&  \frac{2(N_{\eta}+1)^2+1}{6}~,\\
D^{\rm cheb}_{N_{\eta}N_{\eta}} &=& -\frac{2(N_{\eta}+1)^2+1}{6} ~,\\
D^{\rm cheb}_{ij}     &=&  \frac{c_j}{c_j}\frac{(-1)^{i+j}}{x_i-x_j}\,,\quad i,j=2,\,3,\,...,\, N_{\eta}-1\quad (i \ne j) ~,\\
D^{\rm cheb}_{ii}     &=& -\sum_{j=1, j \ne i}^{N_{\eta}} D^{\rm cheb}_{ij} \,, \quad i=2,\,3,\,...,\, N_{\eta}-1 ~.
\label{eq:D_ij}
\end{eqnarray}
Here, $c_i=2$ for $i=1,N_{\eta}$, $c_i=1$ for $1<i<N_{\eta}$. The second-order differential matrix is obtained by doing a matrix multiplication
$D^{\rm cheb,2} \equiv D^{\rm cheb} D^{\rm cheb}$. Thus we have
\begin{equation}
\left.\left(\frac{\partial^2}{\partial \eta^2}+\frac{1}{\eta}\frac{\partial}{\partial \eta}\right)X_{k_j}(\eta,t)\right|_{\eta=\eta_i}=\sum_l \left(D^{\rm cheb,2}_{il}+\frac{1}{\eta_i}D^{\rm cheb}_{il}\right)X_{k_j}(\eta_l,t) ~.
\label{eq:cheb_diff}
\end{equation}
We impose zero boundary conditions along $\eta$, so we only need to solve $X_{k_j}(\eta_i)$ for points $1<i<N_{\eta}$. Correspondingly, only the inner part of the differential matrix is needed. We denote $D^{\rm cheb'}$ and $D^{\rm cheb,2'}$ as the differential matrices excluding the first and last rows and columns. While $D^{\rm cheb}$ is singular, $D^{\rm cheb'}$ is not. Thus the modified differential operator in Eq. (\ref{eq:cheb_diff})
\begin{equation}
P'_{il} \equiv D^{\rm cheb,2'}_{il}+\frac{1}{\eta_i}D^{\rm cheb'}_{il}
\label{eq:cheb_diff_new}
\end{equation}
can be diagonalized as
\begin{equation}
P' = R^{-1} \Lambda R ~, 
\label{eq:cheb_diff_diag}
\end{equation}
where $\Lambda={\rm diag}(-\lambda_1^2, -\lambda_2^2, ..., -\lambda_{N_{\eta}-1}^2)$ with $-\lambda_i^2$ the eigenvalues of matrix $P'$. Using this property, the
Poisson equation (\ref{eq:X_k_nabla}) is solved by
\begin{equation}
\Phi_k = -R^{-1}\,\left(\frac{4 \pi G_N\, m_a}{k^2+\lambda^2} \right) R\,\rho_k ~,
\label{eq:poisson_sol}
\end{equation}
where $\rho_k$ is the Fourier transform of $\rho$ along the $z$ axis. The potential in the real space is obtained by doing an inverse Fourier transform on $\Phi_k$. Having the potential $\Phi$, the potential operator $e^{-i\, W \Delta t}=e^{-i\, \Phi \Delta t}$ in Eq. (\ref{eq:exp_H_M}) is easy to calculate.

Likewise, the kinetic operation $e^{-i\,K\Delta t}$ on the wave function in Eq. (\ref{eq:exp_H_M}) is calculated by
\begin{equation}
e^{-i\, K \Delta t}\psi_k = R^{-1}\,\left[e^{-i\frac{1}{2 m_a}(k^2+\lambda^2)\Delta t}\right] R\,\psi_k ~,
\label{eq:schro_sol}
\end{equation}
where $\psi_k$ is the Fourier transform of the wave function along the $z$ axis. Again, we perform an inverse Fourier transform to obtain the wave function in the real space.

Within one time step, we repeat the above processes several times following the order given by Eq. (\ref{eq:exp_H_M}) with the coefficients given by Eq. (\ref{eq:M_par}).

\begin{figure}[th!]
\begin{center}
\includegraphics[width=.5\textwidth]{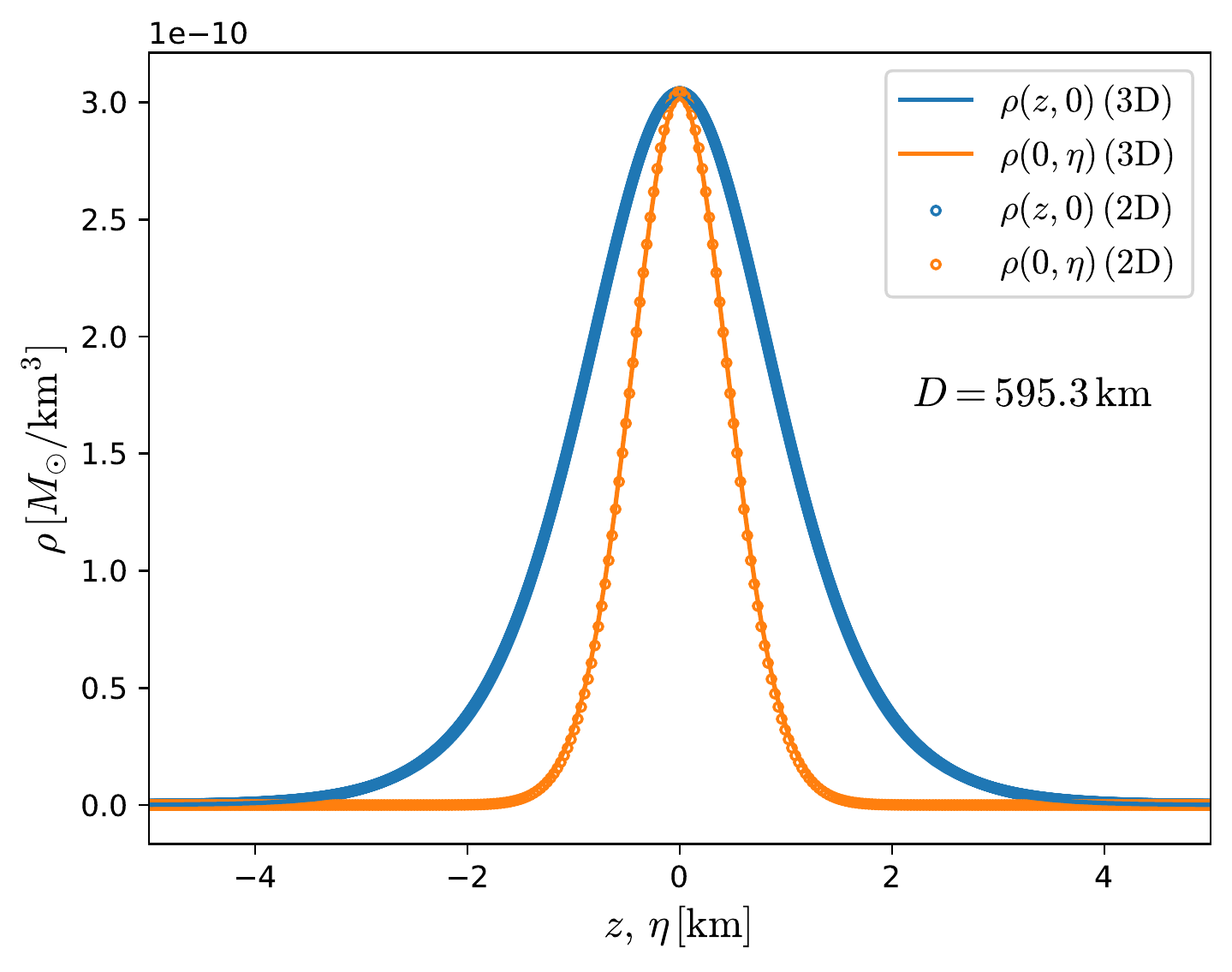}
\end{center}
\caption{Density profiles along $z$ and $\eta$ axis from the 2D head-on collision simulation compared with that from the 3D simulation. The initial condition is the same as in Fig.~\ref{fig:slice} with the axion star mass $M_{a \odot}=10^{-9} M_{\odot}$ and the axion particle mass $m_a=10^{-5}{\rm eV}$.}
\label{fig:3D_2D}
\end{figure}

In Fig.~\ref{fig:3D_2D}, we show a comparison of the results from the new method described above with that from a 3D simulation in Cartesian coordinates. Within the resolution limit of the 3D simulation, we find excellent agreements. On the other hand, the 2D simulation is faster and has much less memory cost enabling us to simulate the axion star to a closer distance to the neutron star.

\section{Free-fall particle approximation}
\label{sec:approximation}
In this appendix, we collect formulae in the free-fall particle approximation.
The trajectory of an axion particle under the external gravitational field sourced by a neutron star sitting at the origin is analytically solved.
Given initial conditions at $t=0$,
\begin{align}
&&\left.z\right|_{t=0}=z_i,
&&\left.\eta\right|_{t=0}=\eta_i,
&&\left.\dot{z}\right|_{\text{initial}}=v_{zi},
&&\left.\dot{\eta}\right|_{\text{initial}}=v_{\eta i},
\end{align}
the solution is
\begin{align}\label{Eq:modified_trajectory}
&m_a t=\beta\left(\frac{m_a}{2E}\right)^{3/2}\bigg[(e\sinh\xi-\xi)-(e\sinh\xi_i-\xi_i)\bigg]\,,
&&r=\sqrt{z^2+\eta^2}= \frac{\beta}{2E}(e\cosh\xi-1) ~,
\end{align}
 \begin{align}
 &m_a \begin{pmatrix}
z \\ \eta
 \end{pmatrix}
 =\sqrt{\frac{m_a}{2E}}
\mathcal{M}_\varphi
 \begin{pmatrix}
\sqrt{\dfrac{m_a}{2E}} \,\beta\,(e-\cosh\xi) \\ M\sinh\xi
 \end{pmatrix},
 &&\mathcal{M}_\varphi=
   \begin{pmatrix}
  \cos \varphi & \sin \varphi  \\
  -\sin \varphi & \cos \varphi 
   \end{pmatrix},
\end{align}
where $\beta=G_N M_{\rm NS}\,m_a$ is the strength of the gravitational interaction; $\xi$ is the parameter which moves from $\xi_i$ (initial time) to $\infty$; $(E, M)$ are integration constants; $e=\sqrt{1+\frac{2EM^2}{m_a\beta^2}}$ is the eccentricity; and 
{\small
\begin{align}
&\cos \varphi = \frac{1}{e}\left(\frac{M v_{\eta i} \left(1+\frac{2 E \,r_{i}}{\beta}\right)}{m_a r_{i} v_{i}^2}+\sqrt{\frac{1-\frac{M^2}{r_{i}^2 m_a^2v_{i}^2}}{1+\frac{v_{\eta i}^2}{v_{zi}^2}}}\right)\,,
&&\sin \varphi = \frac{1}{e}\left(\frac{M v_{zi} \left(1+\frac{2 E \,r_{i}}{\beta}\right)}{m_a r_{i} v_{i}^2}-\frac{v_{\eta i}}{v_i}\sqrt{1-\frac{M^2}{r_{i}^2 m_a^2v_{i}^2}}\right) ~, \nonumber 
\\
&E=\frac{1}{2}m_a v_i^2 - \frac{\beta}{r_i} \,,
&&M=  m_a \left(v_{\eta i} \,z_i - v_{zi}\, \eta_i\right) ~. \nonumber
\end{align}
}
Here we have defined $v_i=\sqrt{v_{yi}^2+v_{zi}^2}$ and $r_i=\sqrt{z_i^2+\eta_i^2}$.

The velocity of an axion particle is given by
 \begin{align}
 &\begin{pmatrix}
\dot{z} \\ \dot{\eta}
 \end{pmatrix}
 =\frac{1}{m_a\,r}
\mathcal{M}_\varphi
 \begin{pmatrix}
-\sqrt{\dfrac{m_a}{2E}} \,\beta \,\sinh\xi \\ M \,\cosh\xi 
 \end{pmatrix},
 &&\dot{r} = \sqrt{\dfrac{2E}{m_a}}\frac{e \cosh\xi }{e\cosh\xi-1}  
 = \sqrt{\frac{2E}{m_a}}\left( 1+ \frac{2E \,r}{\beta} \right) ~.
\end{align}

The density profile before the shell crossing can be computed under the free-fall particle approximation.
Let us consider the time evolution from the initial time $t=0$ to a later time $t=t_1$.
We consider four particles placed at $(\eta_0,\theta_0, z_0)$, $\left(\eta_0\pm\frac{d\eta}{2},\theta_0, z_0\right)$, $\left(\eta_1\pm \frac{a_{\eta\eta}}{2} d\eta,\theta_0, z_1\pm\frac{a_{\eta z}}{2}d\eta\right)$, and $\left(\eta_0,\theta_0, z_0\pm\frac{dz}{2}\right)$. 
At $t=t_1$, the position of these particles is parametrized by $(a_{\eta\eta}, a_{\eta z}, a_{z\eta}, a_{zz})$ as follows.
\begin{align}
&(\eta_0,\theta_0, z_0)
&&\Rightarrow
&&(\eta_1,\theta_0,z_1) ~,
\nonumber\\
&\left(\eta_0\pm\frac{d\eta}{2},\theta_0, z_0\right)
&&\Rightarrow
&&\left(\eta_1\pm \frac{a_{\eta\eta}}{2} d\eta,\theta_0, z_1\pm\frac{a_{\eta z}}{2}d\eta\right) ~,
\nonumber\\
&\left(\eta_0,\theta_0\pm\frac{d\theta}{2}, z_0\right)
&&\Rightarrow
&&\left(\eta_1, \theta_0\pm\frac{d\theta}{2}, z_1 \right) ~,
\nonumber\\
&\left(\eta_0,\theta_0, z_0\pm\frac{dz}{2}\right)
&&\Rightarrow
&&\left(\eta_1\pm\frac{a_{z\eta}}{2} dz, \theta_0, z_1\pm\frac{a_{z z}}{2}dz\right) ~,
\end{align}
where $(\eta_{0\,(1)},\theta_{0\,(1)},z_{0\,(1)})$ is the position at $t=0\,(t_1)$, $d\eta$ and $d\theta$ are small quantities, and $\theta$ is time-independent thanks to the axial symmetry.
The change of the infinitesimal volume is
\begin{align}
&\eta_0  \,d\eta \,d\theta\, dz
&&\Rightarrow
&&\eta_1\,  J\, d\eta\,  d\theta\, dz \,,
&& J:= \left|\det\begin{pmatrix} 
a_{\eta\eta} & a_{\eta z}\\
a_{z\eta} & a_{zz}\\
\end{pmatrix}\right| ~.
\end{align}
From the conservation of the mass, the density change is
\begin{align}
\eta_1 \,\rho_1 (\eta_1,  z_1)=\frac{\eta_0 \, \rho_0 (\eta_0, z_0)}{J} ~,
\end{align}
where $\rho_0$ and $\rho_1$ are density profiles at $t=0$ and $t_1$, respectively.

\setlength{\bibsep}{3pt}
\bibliographystyle{JHEP}
\bibliography{main}

 \end{document}